\documentclass[reprint,superscriptaddress,--
amsmath,amssymb,aps]{revtex4-2}

\usepackage{graphicx}
\usepackage{dcolumn}
\usepackage{bm}
\usepackage{color}
\usepackage[pdfusetitle,%
            bookmarks=true,%
            colorlinks,%
            linkcolor=blue,%
            citecolor=blue,%
            urlcolor=blue%
            ]{hyperref}
\usepackage{physics}

\begin{document}

\title{Phase diagram of one-dimensional deconfined quantum criticality with phonons}

\title{Tensor network study of deconfined quantum criticality in a one-dimensional spin-phonon model}

\newcommand{\TUM}{\affiliation{Technical University of Munich, TUM School of Natural Sciences, Physics Department, 85748 Garching, Germany}}
\newcommand{\MCQST}{\affiliation{Munich Center for Quantum Science and Technology (MCQST), Schellingstr. 4, 80799 M{\"u}nchen, Germany}}
\newcommand{\PKS}{\affiliation{Max-Planck-Institut f\"{u}r Physik komplexer Systeme, N\"othnitzer Straße 38, 01187 Dresden, Germany}}
\newcommand{\USD}{\affiliation{Center for Quantum Mathematics, University of Southern Denmark, Campusvej 55, 5230 Odense, Denmark}}

\author{Anton Romen} \TUM \MCQST
\author{Josef Willsher} \PKS
\author{David Hofmeier} \USD
\author{Johannes Knolle} \TUM \MCQST
\author{Michael Knap} \TUM \MCQST

\date{\today}

\begin{abstract}
{Deconfined quantum critical points (DQCPs) describe continuous transitions between ordered phases beyond the Landau paradigm. A simple example is the N\'eel antiferromagnet (AFM) to valence bond solid (VBS) transition in a 1D antiferromagnetic $J_1-J_2$ model. In analogy to the spin-Peierls instability of critical spin chains, DQCPs are predicted to be unstable towards lattice distortions below a critical phonon frequency. In this work, we use tensor network simulations to investigate this instability in the antiferromagnetic $J_1-J_2$ model coupled to lattice vibrations. We confirm the stability of DQCP for large phonon frequencies and demonstrate that the transition turns strongly first-order below a critical frequency. The instability is caused by a reduction of the Luttinger parameter due to spin-phonon interactions and we identify the effective theory of the behavior as the double sine-Gordon model. The same effective theory is known to describe the classical Ashkin-Teller model, which enables us to show that the critical endpoint is in the four-state Potts universality class. We argue based on non-linear bosonization that the sign of the double-frequency term is fixed, which provides a microscopic justification for the first-order transition. Furthermore, we provide quantitative numerical scaling results for the phonon spectral function, offering an experimental signature to probe DQCP-phonon coupling in low-dimensional materials.} 
\end{abstract}

\maketitle

\section{\label{sec:intro}Introduction}

Deconfined quantum criticality (DQC) was originally introduced \cite{Senthil2004, Senthil2004prb} as a non fine-tuned continuous transition between an antiferromagnetic N\'eel state (AFM) and a dimerized valence bond solid (VBS) in a frustrated magnet on the square lattice. This should be contrasted with Landau-Ginzburg theory in which such direct transitions between phases that break different symmetries are fine-tuned  \cite{Landau2013, Wilson1974}. More generally, DQC has been proposed as a powerful tool to organize competing phases and their transitions in various contexts, including \cite{Liu2019, Khalaf2021, Christos2020, Christos2024, Liu2022, Zhang2020, Nikolaenko2023, Gotz2024, Liu2025, Chen2026, Weber2025, Lou2007}. The scenario of DQC emerges from topological defects of one ordered phase that carry the charge of the other phase \cite{Levin2004,Senthil2006}. These excitations admit a description in terms of fractionalized spinons coupled to emergent gauge fields, which become deconfined at the critical point. Additionally, emergent continuous symmetries at the transition are believed to be a defining feature of these exotic phase transitions \cite{Nahum2015,Wang2017}.

\begin{figure*}
    \centering
    \includegraphics[width=\textwidth, trim=10 10 10 10]{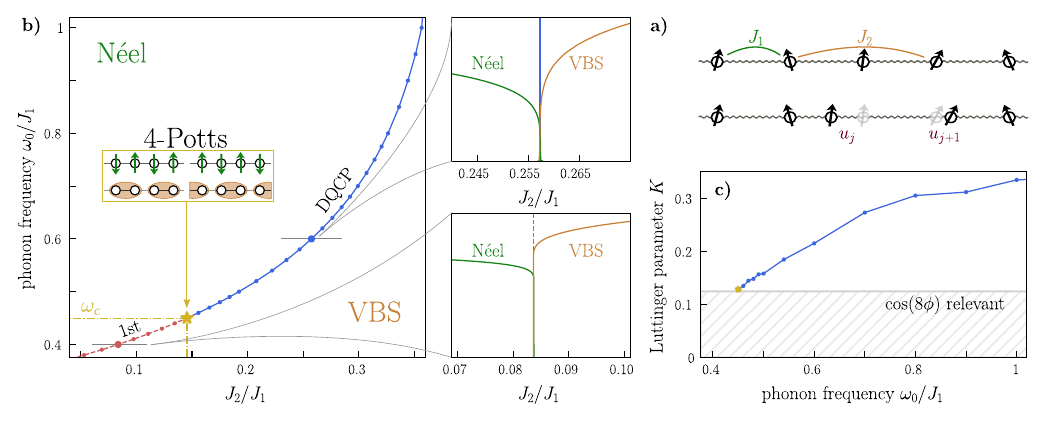}
    \caption{\textbf{Phase Diagram of a frustrated 1D spin-chain coupled to quantum phonons: } a) We consider a 1D spin chain with antiferromagnetic nearest and next-nearest neighbor interactions and strong spin anisotropy $\Delta=4$. Spins are coupled to lattice distortions, modeled by site phonons with displacement $u_j$ at a single variable frequency $\omega_0$. b) Above the critical frequency $\omega_c$, the spin-phonon model realizes a line of deconfined quantum critical points with varying critical exponents between an Ising-N\'eel ordered phase and a dimerized valence bond solid. The critical line ends at $\omega_c$, marked by an endpoint in the four-state Potts universality class. Below $\omega_c$ the transition turns strongly first-order. Typical cuts at fixed $\omega_0$ show differing behavior of the order parameters. At the first order transition, the order parameters jumps discontinuously at the transition. Conversely, along the line of DQCPs, the order parameters continuously vanish at the critical point. c) The dominant effect of dynamical phonons is a reduction of the Luttinger parameter $K$ as function of frequency. The endpoint of the critical line is marked by a Luttinger parameter $K=1/8$, where the next symmetry-allowed term $\cos (8\phi)$ becomes relevant in the continuum action.}
    \label{fig:phase_diagram}
\end{figure*}   

In 1D, continuous order-to-order transitions can arise from an unstable Luttinger liquid \cite{Giamarchi2003}, prominent examples being the AFM-VBS transition in a 1D frustrated $J_1-J_2$ model \cite{Haldane1982} and the Bond-Order Wave to Charge-Density Wave transition in extended 1D Hubbard models \cite{Nakamura2000,Sandvik2004}. The underlying mechanism was already introduced by Haldane \cite{Haldane1982}: Depending on the sign of the umklapp process, a frustrated $J_1-J_2$ Heisenberg chain can be driven into an AFM or VBS ground state. At the boundary between the two phases, the umklapp process vanishes and the system realizes a gapless Luttinger liquid with varying critical exponents. More recent work \cite{Jiang2019, Roberts2019, Huang2019} has proposed this setting as a 1D analogue of DQC, noting that the same effective theory may emerge from microscopic models only after an appropriate (fractionalized) slave-particle description, wherein the order parameters are related to fractionalized degrees of freedom in close analogy to the 2D case. The AFM-VBS transition in the $J_1-J_2$ spin chain with strong easy-axis anisotropy \cite{Mudry2019, Lee2023} was revisited as a particularly simple example, wherein the low-energy effective theory takes the form of a sine-Gordon model \cite{Giamarchi2003} with emergent $U(1)$ symmetry at the critical point \cite{Lee2023}. The same long-wavelength theory also arises in other microscopic realizations of 1D DQC, including ~\cite{Yang2020,Roberts2021,Yang2023,Baldelli2024,Romen2024}.

It is well known that a spin-1/2-chain realizing a gapless phase, i.e., the Luttinger liquid, is unstable against distortions of the underlying lattice \cite{Giamarchi2003, Pytte1974, Peierls1955, Cross1979}, by a mechanism called the spin-Peierls instability \cite{Pincus1971,Pytte1974}. The same mechanism has also recently been discussed in the context of critical 2D spin liquids \cite{seifert2024, Ferrari2025}. A simple energetic argument shows that the chain can lower its energy by spontaneously dimerizing and forming singlet states, thereby realizing the gapped VBS phase. The same reasoning can also be extended to a gapless phase transition, like the aforementioned DQCP. Therefore, one may reasonably suspect that the 1D DQCP is generically unstable with the transition turning first-order when taking into account lattice degrees of freedom. Indeed, in a recent work \cite{Hofmeier2024} it was argued, based on a low-energy field theoretical framework, that an analog of the spin-Peierls instability is also expected for DQC in the case of coupling to phonons: Static distortions of the underlying lattice geometry turn the transition strongly first order. This instability is potentially lifted upon adding dynamics of the underlying lattice; considering dynamical phonon excitations of the lattice, it has been argued that DQC remains stable above a critical phonon frequency~\cite{Hofmeier2024}. 

In this context, recent numerical studies on 2D systems indeed suggest that DQC and critical quantum spin liquids show confining behavior as a function of phonon frequency \cite{gotz2024tuning,Ferrari2025}. Conversely, in 1D it is known that a continuous transition can turn strongly first order below a critical value of the Luttinger parameter, e.g. \cite{Sandvik2004,Yang2023}. Howerver, the effect of lattice coupling remains largely unexplored beyond the Heisenberg point \cite{Sandvik1999,orignac2004,citro2005}. This highlights that further work is needed to understand the fate of deconfined phase transitions and emergent symmetries under the presence of phonon coupling. In this work, we aim to understand the mechanisms of such a deconfined phase transition turning strongly first order when coupled to lattice degrees of freedom by focusing on a one-dimensional model, allowing us to perform both analytic and numerical analyses. With this complementary insight we are able to understand the nature of the DQC-endpoint and predict the consequences of the low-frequency confining phase.

\subsection{Summary of main results}

We study a prototypical 1D spin-chain coupled to phonons and determine the full phase diagram of this model shown in Fig.~\ref{fig:phase_diagram} using tensor network simulations by implementing the phonons as truncated bosonic sites. Our numerics is consistent with the DQCP persisting above a critical phonon frequency $\omega_c$. We further extract the dynamically varying Luttinger parameter $K$ which enables us to relate the breakdown of DQC to a reduction of the Luttinger parameter resulting from spin-phonon coupling.  

We argue that the critical endpoint of the continuous DQCP lies in the four-state Potts universality class and is described by the same effective double-frequency sine-Gordon theory as the high-symmetry critical point of the Ashkin-Teller model \cite{Ashkin1943,Kohmoto1981,Aoun2024}. Contrary to the Ashkin-Teller model that develops an intermediate phase beyond the endpoint, our model realizes a strong first-order transition. Based on a classical analysis of the double-frequency sine-Gordon model, we show that the nature of the transition depends on the prefactor $\rho$ of the double frequency $-\rho\cos(8\phi)$ term: In the Ashkin-Teller model, the sign of $\rho$ is fixed to be \textit{negative}, which causes the line of continuous transitions to split into two Ising transitions. By contrast, a \textit{positive} prefactor introduces a potential barrier at the critical point when the $\cos(8\phi)$ term is relevant, turning the DQCP strongly first order. We show that the sign of $\rho$ is indeed fixed to be \textit{positive} in non-linear bosonization, thereby providing a microscopic justification of the spin-Peierls instability of 1D DQCP.

Our work further provides insight into the physical effects of spin-phonon coupling by computing the phonon spectral function across the continuous and first-order transition. We study how spin fluctuations modify the spectral properties of the phonons. A famous example of this is the Kohn anomaly of low-dimensional electron systems which is a discontinuity of the phonon dispersion due to (critical) fluctuations in a coupled electron system \cite{kohn1959}. The Kohn anomalies in the free electron chain and in (interacting) spin chains provide another perspective on the (spin-)Peierls instability: Increasing the spin/electron-phonon coupling acts to renormalize the phonon self energy, until it eventually condenses when the gap closes. 
In our numerics, we indeed observe a Kohn anomaly in the phonon dispersion across the DQCP. This manifests as a kink in the phonon spectrum in the ordered phases, as well as the emergence of a critical continuum at low frequencies at the deconfined transition. We measure the phonon spectral function $A(q,\omega)$ and find that at the critical point, spin-phonon coupling leads to a hybridization with the dimer-fluctuations \cite{luther1974}, such that a universal power-law response emerges $A(q=\pi,\omega)\sim\omega^{-2K}$, governed by the Luttinger parameter of the transition.

\subsection{Structure of the work}

The work is organized as follows: In Sec.~\ref{sec:DQCP-and-static} we introduce the spin-phonon model and discuss various limiting cases. First, we analyze the static limit of the decoupled spin chain in Sec.~\ref{sec:DQCP} that realizes a DQCP between an AFM and VBS phase and give a brief summary on DQCP in (1+1)D. We then establish in Sec.~\ref{sec:DQCP-phonon-coupling} how spin-phonon coupling naturally emerges in this model and introduce suitable approximations for the phonon and spin-phonon Hamiltonians. In Sec. \ref{sec:Spin-Peierls} we consider the adiabatic limit of the coupled spin-phonon model, akin to static lattice distortions. In this limit, lattice vibrations can be treated classically, resulting in a static spin Hamiltonian with dimerized nearest-neighbor interactions and a quadratic potential energy cost. 
We use this model to give a brief overview of the well known Peierls instability and recapitulate predictions from Ref.~\cite{Hofmeier2024} that the DQCP turns strongly first-order on a distorted lattice. 
We provide numerical data supporting this scenario. In Sec. \ref{sec:model-phase-diagram} we extend the analysis to the general case of dynamical spin-phonon coupling. We determine the phase diagram of the model and link the nature of the transition to the flow of the Luttinger parameter as summarized in Fig.~\ref{fig:phase_diagram}. We provide an analysis of the critical endpoint of the DQC line and link the low-energy theory of the model to the static limit. In Sec. \ref{sec:phonon-response}, we  analyze the phonon spectral function, and find numerical evidence for the emergence of a critical continuum in the second-order regime and large weight at small frequencies in the other regime, consistent with a 1st order transition. This also provides a powerful observable to investigate spin-phonon coupling in scattering experiments. We provide a summary of results and outlook for future work in Sec.\ref{sec:discussion}.
Technical details are relegated to the appendices.

\section{Spin-phonon model: limiting cases}
\label{sec:DQCP-and-static}

We consider a one-dimensional, coupled spin-lattice system
\begin{equation}
    H = H_s + H_p + H_{sp}\,,
    \label{eq:H-total}
\end{equation}
where $H_s$ describes the spin Hamiltonian, $H_p$ the phonon Hamiltonian of lattice vibrations, and $H_{sp}$ the associated spin-phonon coupling. We will introduce the exact forms below in Eqs.~\eqref{eq:J1J2XXZ_und}, \eqref{eq:HPhonons} and \eqref{eq:HSpinPhonons}. 

We  introduce the model in steps by looking into various limiting cases. First, we  discuss the decoupled spin model $H_s$, corresponding to the limit of a \emph{regular lattice} in a phonon-free model and demonstrate that it realizes a DQCP between a N\'eel antiferromagnet and valence bond solid, similar to Ref.~\cite{Mudry2019}. In this context, we will also introduce the numerical method used to study the spin-phonon model and the lattice and continuum theories of DQCP in 1D.
We then proceed with a general discussion on spin-lattice coupling and introduce models for $H_p$ and $H_{sp}$. After that, we investigate the adiabatic limit, corresponding to lattice vibrations of infinitesimal energy cost. The result is a staggered spin Hamiltonian of \textit{static} lattice distortions, which we use to illustrate an analog of the spin-Peierls instability for DQCP. The fully dynamical spin-phonon model will be investigated in the next section.

\subsection{Phonon-free limit: Deconfined quantum criticality}
\label{sec:DQCP}

\subsubsection{Spin Hamiltonian}

The spin-part of our model is defined by the following spin-1/2 chain
\begin{eqnarray}
    \begin{aligned}
        H_{s} = J_1&\sum_j[\mathbf{S}_j \cdot \mathbf{S}_{j+1} + (\Delta_1-1)S^z_j S^z_{j+1}] \\
 +        J_2&\sum_j[\mathbf{S}_j \cdot \mathbf{S}_{j+2} + (\Delta_2-1)S^z_j S^z_{j+2}]    
    \end{aligned}
    \label{eq:J1J2XXZ_und}
\end{eqnarray}
 with nearest neighbor $J_1$ and next-nearest neighbor $J_2$ exchange couplings. The parameters $\Delta_{1,2}$ are XXZ anisotropy parameters, which we  take to be equal $\Delta\equiv\Delta_1=\Delta_2$ unless specified otherwise. The regime of interest for this work will be $\Delta > 1$, where it is well known that the $J_2=0$ limit realizes an antiferromagnet with N\'eel order (AFM). Increasing $J_2/J_1$ introduces frustration into the AFM ground state and drives the model towards a valence-bond-solid (VBS) ground state, consisting of singlets on every second bond.
 
\subsubsection{Continuum field theory}
We can describe the long-wavelength behavior of the model introduced in Eq.~\eqref{eq:J1J2XXZ_und} as a sine-Gordon model by using a standard Jordan-Wigner transformation and bosonization \cite{Giamarchi2003}, as briefly outlined in Appendix \ref{a:boso}. It produces a low-energy description in terms of a compact bosonic field $\phi(x) \in [0,2\pi]$, with the action
\begin{equation}
        S_{\mathrm{SG}} = \int \dd{x}\dd{\tau} \left[\frac{1}{2\pi K}\left((\partial_x\phi)^2+(\partial_\tau\phi)^2\right) 
    - \mu\cos(4\phi)\right] \,.
    \label{eq:SG}
\end{equation}
The parameter $K$ is the Luttinger parameter, whose precise value is determined by the microscopic couplings $J_1, J_2$ and $\Delta$. The scaling dimension of the coupling $\mu$ is determined by the Luttinger parameter and given by $[\mu] = 4K$; therefore, the interaction $\cos(4\phi)$ is \emph{irrelevant} ($[\mu]>2$) when $K>1/2$.
The value of the coupling $\mu \sim 2J_1\Delta-2J_2(2+\Delta)$ is likewise determined by the parameters of the microscopic theory; for the nearest-neighbor only model ($J_2=0$) the interaction is irrelevant for $\Delta \leq 1$.
This theory flows to a conformal fixed point with an emergent U(1) symmetry, called the Luttinger liquid phase of the XXZ chain \cite{Giamarchi2003}. In this regime, the sine-Gordon model in Eq.~\eqref{eq:SG} flows in the infra-red (long-distance limit) to the compact boson CFT with central charge $c=1$ and compactification radius determined by the Luttinger parameter $K$. 

If instead we consider theories with XXZ anisotropy $\Delta > 1$, we find that the interaction $\cos(4\phi)$ is now \emph{relevant}. The result is a gapped phase, with order determined by the sign of $\mu$. The competing order parameters of the spin model in Eq.~\eqref{eq:J1J2XXZ_und} can be written directly in terms of the bosonic continuum field $\phi(x_j)$. We are interested in the N\'eel ($M^z_j$) and valence-bond ($\Psi_j$) order parameters, defined as
\begin{eqnarray}
    \begin{aligned}
    M^z_j &\equiv (-1)^{j}S^z_j \sim \cos(2\phi(x_j))\,, \\ 
        \Psi_j &\equiv \mathbf S_j \mathbf S_{j+1} - \mathbf S_{j+1} \mathbf S_{j+2} \sim \sin(2\phi(x_j)).
    \end{aligned}
\label{eq:op-definition}
\end{eqnarray}
Using these mappings, we may rewrite the interaction in Eq.~\eqref{eq:SG} as
\begin{equation}
  -\mu\cos(4\phi) = \mu \left[\sin^2(2\phi) -  \cos^2(2\phi) \right]
   = \mu(\Psi^2-M_z^2) \, ,
\end{equation}
such that N\'eel order is preferred when $\mu>0$, and dimer order when $\mu<0$.
For the nearest-neighbor model ($J_2=0$), bosonization predicts $\mu>0$ and we recover the gapped Ising-N\'eel ground state. Increasing $J_2/J_1$ introduces frustration and drives the model towards a dimerized phase. By tuning appropriate parameters in the microscopic Hamiltonian, it is possible to cause a sign change in $\mu$ and drive a transition to valence bond order.
The critical point $\mu=0$ is described by the same sine-Gordon continuum action as the Luttinger liquid, but with a Luttinger parameter $K<1/2$.
The gap opens with exponent $|\mu|^{\nu_{\mu}}$, where $\nu_\mu^{-1} = 2-4K$.
The order parameters onset with power-law behavior $M^z,\Psi \sim |\mu|^{\beta}$ governed by the exponent $\beta = \nu_\mu K$. Correlations of the order parameters at the transition can be obtained from Eq.~\eqref{eq:op-definition} using standard bosonization techniques \cite{Giamarchi2003}:
\begin{equation}
    \langle M^z(0)M^z(r) \rangle  \sim \langle e^{2i\phi(0)\phi(r)} \rangle \sim r^{-2K}
    \label{eq:op-correlations}
\end{equation}
and similarly $\langle \Psi(0)\Psi(r)\rangle \sim r^{-2K}$. 

Next, we briefly comment on the symmetries of the Hamiltonian in Eq.~\eqref{eq:J1J2XXZ_und}. The model is invariant under lattice translations $\mathbf{T}_x$ and possesses a $\mathbb{Z}_2 \times U(1)$ spin symmetry, corresponding to a global spin flip along the z-axis and continuous rotations in the xy-plane. Under these symmetries, the field $\phi$ transforms as
\begin{eqnarray*}
    \begin{aligned}
        \phi &\xrightarrow{\mathbb{Z}_2\hspace{3mm}}-\phi+\pi/2  \,,\\
        \phi &\xrightarrow{U(1)}\phi \,,\\
        \phi &\xrightarrow{\mathbf{T}_x\hspace{3mm}}\phi+\pi/2\,.
    \end{aligned}
\end{eqnarray*}
For Luttinger parameter $K>1/8$, the only symmetry-allowed relevant perturbation is of the form $\cos(4\phi)$. Consequently, the sine-Gordon action in \eqref{eq:SG} constitutes the most general low-energy effective theory consistent with the symmetries of the model.
\begin{figure}
    \centering
    \includegraphics{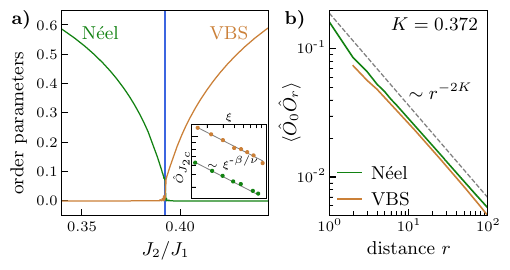}
    \caption{\textbf{Summary of Deconfined Criticality in a 1D spin chain:} a) Upon tuning $J_2/J_1$, the model undergoes a continuous transition from a gapped phase with N\'eel order to a dimerized VBS phase. Numerical data displays a weakly first-order transition with length scale set by the MPS correlation length $\xi$. Upon increasing the bond dimension $\chi$, order parameter residuals at the transition vanish as $\hat O_{J_{2c}}\sim \xi(\chi)^{-\beta/\nu}\sim \xi^{-K}$ (inset), indicating a continuous transition in the thermodynamic limit, see main text for details. b) Order parameter correlations at the transitions decay algebraically, with exponent set by the Luttinger parameter $K$ of the critical theory.}
    \label{fig:DQCP_static}
\end{figure}

\subsubsection{Numerical method}

Throughout this work we simulate the model introduced in Eq.~\eqref{eq:H-total} using infinite-system Density Matrix Renormalization Group (iDMRG) simulations \cite{White1992, Schollwock2011} implemented within the TeNPy library \cite{Hauschild2024} and optimize over infinite Matrix Product States (iMPS) in the thermodynamic limit \cite{Vidal2003, Vidal2007}. We provide further details on the numerical implementation in Appendix \ref{A:numerics}.

Let us start by considering the spin Hamiltonian $H_s$ in Eq.~\eqref{eq:J1J2XXZ_und} at fixed $\Delta=4$ while varying $J_2/J_1$. As outlined before, we expect the ground state to have AFM order for small $J_2/J_1$ and VBS order at large $J_2/J_1$, which can be captured by the order parameters introduced in Eq.~\eqref{eq:op-definition}. We show in Fig.~\ref{fig:DQCP_static} that the N\'eel order is suppressed upon increasing $J_2/J_1$ until the critical point  $J_{2,c}/J_1=0.3923$, where valence bond order onsets with the same exponent. We additionally measure the critical correlations of both N\'eel and valence bond order parameters introduced in Eq.~\eqref{eq:op-correlations} at the transition and find good agreement with a power-law described by a Luttinger parameter $K=0.372$. 

Following arguments introduced in Ref.~\cite{Roberts2019}, we confirm that this transition is second order in the thermodynamic limit. For an iMPS, any finite bond dimension $\chi$ introduces an effective length scale $L_\text{eff} \sim \xi(\chi)$ set by the correlation length of the variational ground state. Consequently, the resulting transition can never be truly continuous at finite $\chi$ and the order parameters acquire a finite residual $\hat{O}_{J_{2c}}$ at the critical point. As we expect a second-order transition, the residuals should vanish as a function of systems size $L$ as $\hat O_{J_{2c}} \sim L^{-\beta/\nu_\mu} \sim \xi(\chi)^{-\beta/\nu_\mu} \sim \xi(\chi)^{-K}$ with bond dimension $\chi$. 

In practice, the correlation length of an MPS in the thermodynamic limit is obtained from the transfer matrix \cite{Perez2006,Hauschild2024}. Additionally, the residuals are controlled using a sweep protocol in which the transition is approached from within one of the ordered phases. In each simulation, the previous ground state is used as initial state for the subsequent run. In proximity to the critical point, this procedure induces hysteresis and fixes the type of residual. The method, in fact, provides a surprisingly robust way to compute the relevant Luttinger parameter $K$ \cite{Roberts2019, Romen2024}, see Fig. \ref{fig:DQCP_static}. We explicitly confirm that estimates for $K$ obtained from the order parameter residuals and critical order parameter correlations agree; c.f., inset of Fig.~\ref{fig:DQCP_static} a).

\subsection{Spin-phonon coupling}
\label{sec:DQCP-phonon-coupling}

We now introduce models for $H_p$ and $H_{sp}$ in Eq.~\eqref{eq:H-total} to capture the effect of spin-phonon coupling on the DQCP.

\subsubsection{Phonons}
We start by introducing the distortions of lattice sites around their equilibrium positions $u_j = x_j-\bar x_j$, which we assume to be small. Within the harmonic approximation, $u_j$ may then be modeled by dissipationless, uncoupled Einstein phonons at some frequency $\omega_0$ via a phonon Hamiltonian
\begin{equation}
    H_p =\sum_j \frac{p_j^2}{2m}+\frac{1}{2}m\omega_0^2 u_j^2\,.
    \label{eq:HPhonons}
\end{equation}
By introducing appropriate raising and lowering operators $u_j = 1/\sqrt{(2m\omega_0)}(a+a^\dagger)$,  $H_p$ in second quantized form is given by
\begin{equation}
    H_p = \omega_0\sum_j a^\dagger_ja_j+\text{const.}\,.
     \label{eq:HPhonons-2nd-quant}
\end{equation}
We will use this form to simulate the lattice degrees of freedom as boson-sites truncated at some maximal occupation. We note that instabilities of the DQCP occur at wavevector $q=\pi$, which \emph{a posteriori} justifies considering a single (finite) phonon frequency $\omega_0$ as a good approximation for the (generally flat and gapped) phonon band around the edge of the Brillouin zone $q\approx\pi$.
In a next step, we will study how vibrations of the underlying lattice affect the exchange constants $J_1$ and $J_2$.

\subsubsection{Spin-phonon coupling}

We first note that the origin of Heisenberg-type spin-spin interactions is through exchange interactions of neighboring spinful orbitals. The corresponding exchange constants are naturally bond-length dependent and we may for example model $J_{i,j} \sim J_0 e^{-\tilde{g}|x_i-x_j|}$ in terms of the atomic positions $x_j$. This spatial dependence is used implicitly in Eq.~\eqref{eq:J1J2XXZ_und} in the sense that $J_1 = J_0e^{-a}$, $J_2 = J_0e^{-2a}$, with some lattice constant $a=\bar x_{i+1}-\bar x_i$, where we assumed a regular lattice with equilibrium positions $\bar x_j$. 

Finite displacements ($u_j\ll a$) of the lattice sites render the exchange constants site dependent. To linear order, the nearest neighbor exchange constant $J_1$ acquires a correction
\begin{eqnarray*}
    \begin{aligned}
    J_1 \rightarrow J_{j,j+1} &= J_0e^{-\tilde g |x_j-x_{j+1}|} = J_0e^{-\tilde g(a+u_{j+1}-u_j)} \\ 
    &\approx J_1[1+\tilde g (u_j-u_{j+1})]\,.
    \end{aligned}
\end{eqnarray*}
This motivates the following form for the Heisenberg spin-phonon coupling:
\begin{eqnarray}
    \begin{aligned}
    H_{sp} &= J_1\tilde g\sum_j (u_j-u_{j+1})[\mathbf{S}_j \cdot \mathbf{S}_{j+1}+(\Delta-1)S^z_j S_{j+1}^z]\\
    &= \frac{g}{\sqrt{\omega_0}}\sum_j(a_j+a_j^\dagger-a_{j+1}-a_{j+1}^\dagger)\times\\&\qquad\qquad\qquad[\mathbf{S}_j \cdot \mathbf{S}_{j+1}+(\Delta-1)S^z_j S_{j+1}^z]\,,
    \end{aligned}
    \label{eq:HSpinPhonons}
\end{eqnarray}
where we have introduced $g=J_1\tilde g/\sqrt{2m}$. Here, we neglected modifications of $J_2$, which is justified by the observation that lattice configurations are staggered $u_j=(-1)^ju$, such that the bond length of next-nearest neighbors remains.

\subsection{Adiabatic limit: Spin-Peierls instability}
\label{sec:Spin-Peierls}

We start to build some intuition on the influence of lattice vibrations on 1D DQC by considering the adiabatic limit $\omega_0\rightarrow 0$ in Eqs.~\eqref{eq:HPhonons} and \eqref{eq:HSpinPhonons}, corresponding to a lattice with infinitesimal elastic energy cost. 

We start with the phonon Hamiltonian in Eq.~\eqref{eq:HPhonons-2nd-quant}. For $\omega_0\rightarrow 0$, the average number of phonons $n_p \equiv \langle a^\dagger_j a_j\rangle$ is large for any finite energy density. We can thus take the large-N limit and replace the number operator by its expectation value
\begin{align}
    H_p \rightarrow H_\kappa = \kappa N \delta^2
    \label{eq:H-potential-classical}
\end{align}
to arrive at a harmonic potential describing the energy cost of lattice distortions. Here, we introduce an effective lattice stiffness $\kappa = m\omega_0^2$, $N$ denotes the number of lattice sites, and $\delta \sim n_p$ measures the distortion. 

\begin{figure}
    \centering
    \includegraphics{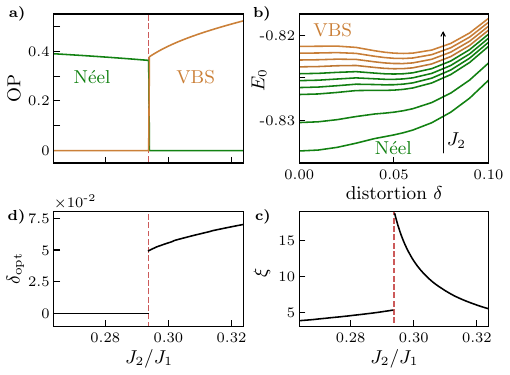}
    \caption{\textbf{Spin-Peierls Instability of DQC in a 1D spin chain:} a) Static lattice distortions drive the DQCP to a strong first-order transition, displayed by a clear jump in the order parameter. b) Inside the VBS phase, the system favours a sudden distortion of the underlying lattice $\delta\neq0$ to lower the ground state energy. c) The lattice instability is captured by a sudden jump of the optimal lattice distortion $\delta_\text{opt}$ at the transition. d) The correlation length displays clear first-order character.}
    \label{fig:1st_order_static}
\end{figure}
\begin{figure*}
    \centering
    \includegraphics[width=0.8\textwidth]{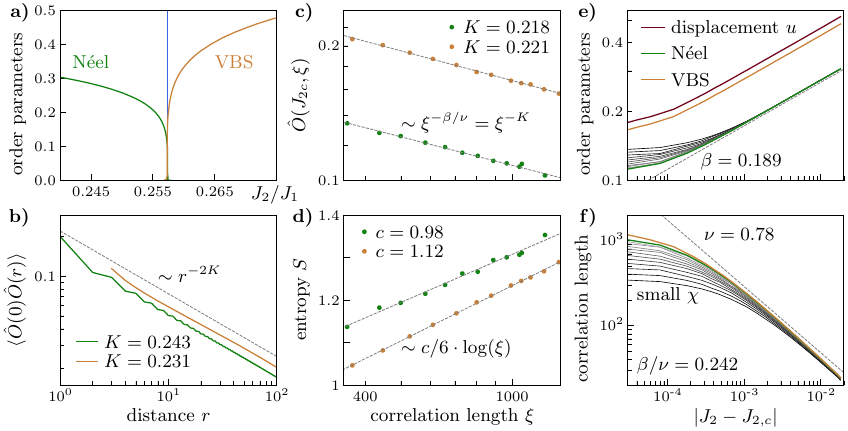}
    \caption{\textbf{DQCP in the spin-phonon model:} a) When coupling to dynamical Einstein phonons with frequency $\omega_0$, the continuous transition between N\'eel and VBS phase persists above a critical phonon frequency $\omega_c$. b) Order parameter correlations at the transition decay algebraically, with power law exponent $-2K$ governed by the Luttinger parameter of the critical theory. c) The order parameters develop a finite residual at the critical point, which vanishes with increasing correlation length $\xi$ of the MPS ground state. The residuals are controlled using a sweep protocol initialized in one of the two ordered phases. Finite-size scaling predicts a power-law decay $\hat O_{J_{2c}} \sim \xi^{-\beta/\nu}=\xi^{-K}$, which we use to obtain an independent estimate for the Luttinger parameter. d) The entanglement entropy at the critical point scales linearly with $\log \xi$, consistent with a free Gaussian fixed point with central charge $c=1$. e) Both spin order parameters as well as the average phonon displacement demonstrate power-law scaling in the vicinity of the transition with matching exponent $\beta$ as required by the critical theory. f) Power-law behavior with increasing bond dimension $\chi$ is also observed in the correlation length and the Luttinger parameter obtained from the ratio of critical exponents is consistent with previously obtained estimates. Data obtained for $\Delta_1=\Delta_2=4, g=0.25J_1, \omega_0 = 0.6J_1$.}
    \label{fig:DQCP_dynamical_overview}
\end{figure*}

Some care has to be taken for the spin-phonon coupling in Eq.~\eqref{eq:HSpinPhonons}. A state with fixed $n_p$ has $\expval{a+a^\dagger} = \expval {u_j}=0$. Instead, we consider coherent states $\ket \alpha$ to account for fluctuations around $n_p$, which are known to be a good approximation when taking the classical limit \cite{Glauber1963}. We thus approximate (assuming $\alpha \in \mathbb{R}$)
\begin{eqnarray*}
    \begin{aligned}
        \delta_j \equiv \expval{u_j} &= 1/\sqrt{2m\omega_0}\bra{\alpha}(a_j+a_j^\dagger)\ket{\alpha}\sim \alpha/\sqrt{m\omega_0} \\
       \bra{\alpha}a^\dagger_ja_j\ket{\alpha} &= \alpha^2 \sim m \omega_0 \delta^2_j\,.       
    \end{aligned}
\end{eqnarray*}
Eq.~\eqref{eq:H-potential-classical} is recovered from the second line of the above equation by inserting the result into Eq.~\eqref{eq:HPhonons-2nd-quant} and summing over lattice sites. 

Lattice distortions can couple to the VBS order parameter with a 2-site unit cell, which motivates considering a staggered lattice distortion $\delta_j = (-1)^j\delta$. This maps $H_{sp} \rightarrow H_\text{dim}$ onto a dimerized spin Hamiltonian 
\begin{align}
    H_{\mathrm{dim}} = &J_1\sum_j\delta(-1)^j[\mathbf{S}_j \cdot \mathbf{S}_{j+1} + (\Delta-1)S^z_j S^z_{j+1}].
\end{align}
To summarize, in the adiabatic limit, the full spin-phonon Hamiltonian in Eq.~\eqref{eq:H-total} is mapped to a static spin Hamiltonian given by
\begin{equation}
H_\text{static}(\delta)=H_s + H_{\mathrm{dim}}(\delta)+H_\kappa (\delta)
\label{eq:H-dqc-static-phonons}
\end{equation}
with $H_s$ as defined in Eq.~\eqref{eq:J1J2XXZ_und}. This is a \emph{static} model of lattice distortions which models the dimerization as a classical field, neglecting temporal fluctuations.
Adding a finite lattice distortion lowers the symmetry of the Hamiltonian and allows for a more relevant operator to be present in the continuum theory \cite{Pincus1971,Pytte1974,Cross1979}
\begin{equation}
    S_{\mathrm{static}}[h\sim\delta] = S_{\mathrm{SG}} + h\int\dd{x}\dd{\tau}\sin(2\phi) \, .
\end{equation}
The scaling dimension of this new operator is $[h]= K$. This is relevant for all $K<2$, and hence strongly relevant at both the Heisenberg point $K=1/2$ as well as in the full DQCP regime $1/8<K<1/2$.
Perturbative arguments suggest that this relevant operator will lead to an energy gain at the DQCP which always outcompetes the cost of lattice distortions \cite{Hofmeier2024}. It implies spontaneous dimerization at the critical point. This has been interpreted as causing a first-order transition, but  remains to be verified numerically.

To measure the competition between the lattice and spin degrees of freedom, we numerically compute the energy of the combined spin-lattice model. This energy, given by Eq.~\eqref{eq:H-dqc-static-phonons} has contributions from the ground state energy of a distorted spin chain, as well as the potential energy cost of the distortion, $\kappa \delta^2$ per site.
To obtain the behavior of the order parameters, we calculate the energy landscape $H_\text{static}(\delta)$ as a function of the distortion parameter. For each $J_2/J_1$, we find the distortion with the lowest energy, and calculate the order parameter of the spin chain at this optimal lattice distortion.

We now present our numerical results which confirm a spin-Peierls instability of the static spin-lattice model.
Tuning $J_2/J_1$ at fixed lattice stiffness $\kappa=3.5$, the order parameters jump at a critical value, consistent with a first order transition; see Fig.~\ref{fig:1st_order_static}a. The landscape of the system's total energy $E_0(\delta)$ as a function of distortion parameter is shown in Fig.~\ref{fig:1st_order_static}b. For low $J_2/J_1$ in the N\'eel phase, the lattice has the lowest energy at zero distortion.
Upon increasing $J_2/J_1$, the energy $E_0(\delta)$ develops a minimum at finite $\delta$; inside the VBS phase, a sudden distortion of the lattice to some $\delta_\text{opt} \neq 0$ lowers the energy of the ground state. 
We obtain the optimal lattice distortion $\delta_\text{opt}$ for each parameter $J_2/J_1$ via a quadratic fit of the energy around the minimum. In summary, we observe a clear jump in both order parameters, the optimal distortion $\delta_\text{opt}$ and a \emph{finite} correlation length at the transition, consistent with a strong first order N\'eel-VBS transition.

\section{Spin-phonon model: Dynamical phonons}
\label{sec:model-phase-diagram}

We now turn to the full Hamiltonian introduced in Eq.~\eqref{eq:H-total}. Integrating out the dynamical phonon field generates a sine-Gordon action
\begin{equation}
        S_{\mathrm{SG}'} = \int \dd{x}\dd{\tau} \left[\frac{1}{2\pi K'}\left((\partial_x\phi)^2+(\partial_\tau\phi)^2\right) 
    - \mu'\cos(4\phi)\right] \,.
    \label{eq:SG_new}
\end{equation}
with renormalized parameters $K'(\omega_0)$ and $\mu'(\omega_0)$, that depend on the phonon frequency $\omega_0$. We reproduce the calculation from Ref.~\cite{Hofmeier2024} in Appendix~\ref{A:SG_phonons}, and state the form of the new parameters as a function of $\omega_0$. 
Importantly, it was shown that the effect of decreasing $\omega_0$ favors the dimer phase, and reduces the parameter $K'(\omega_0) < K$.
Furthermore, it was hypothesized that this theory breaks down at the point $K'(\omega_c)=1/8$, which defines a critical phonon frequency $\omega_c$ below which the DQCP picture breaks down. It is natural that the transition becomes first order below this point, as we just showed to be the case in the static phonon limit $\omega_0\to0$ .

We now investigate the full spin-phonon model with $\Delta_1=\Delta_2=4, g=0.25J_1$. To simplify the electron-phonon coupling, we consider isotropic spin phonon interactions, $H_{sp} \to g\sum_j(a_j+a_j^\dagger-a_{j+1}-a_{j+1}^\dagger)\mathbf{S}_j \cdot \mathbf{S}_{j+1}$, \textit{c.f.}, Eq.~\eqref{eq:HSpinPhonons}. We found this isotropic coupling to be numerically more favorable, while not changing the essential physics. 
We limit the maximum number of phonons per site to $N_\text{max}=6$ and confirm in Appendix \ref{A:data} that the relevant physics does not change with increasing $N_\text{max}$. With that, we investigate the transition by tuning $J_2/J_1$ for various values of $\omega_0/J_1$. For our iMPS, we consider bond dimensions $136 \leq \chi \leq 472$. Further details on the numerical implementation are provided in Appendix \ref{A:numerics}.

\subsection{Results of the 2nd- and 1st-order transition}

In the limit $\omega_0 \to \infty$, the lattice becomes infinitely stiff, causing the spins and phonons to effectively decouple. In this limit, the spin model realizes the DQCP introduced in Fig.~\ref{fig:DQCP_static}. We now show that our numerics is consistent with a continuous transition for finite $\omega_0>\omega_c\approx0.45 J_1$.

We summarize results for the DQCP regime in Fig.~\ref{fig:DQCP_dynamical_overview}, using a typical cut at phonon frequency $\omega_0=0.6J_1$. We first determine the critical point $J_{2c}(\omega_0=0.6J_1)\approx0.2573 J_1$, at which both order parameters continuously vanish, see Fig.~\ref{fig:DQCP_dynamical_overview}a.
The spin-phonon coupling acts to favor the VBS phase, shifting the critical point to lower $J_2/J_1$ compared to the static case.
The order parameter correlations at the transition show clear power-law decay; see Fig~\ref{fig:DQCP_dynamical_overview}b. 
The numerically determined exponent $-2K'_c$, indicates a Luttinger parameter $K'_c \approx 0.237$, at the new critical point, where we use the subscript $K'_c$ to emphasize that the Luttinger parameter was obtained from critical order parameter correlations. Crucially, we find a lower value of $K'_c< K = 0.372$ than in the static limit. This agrees with the prediction that spin-phonon coupling lowers the Luttinger parameter at the transition.

We further compute the order parameter residuals at the transition for various bond dimensions and confirm that the residuals agree with the finite-size scaling prediction introduced in Sec. \ref{sec:DQCP}, as shown in Fig~\ref{fig:DQCP_dynamical_overview}c. 
We thus obtain a second estimate $K'_r(\omega_0=0.6J_1) \approx 0.221$ for the Luttinger parameter of the transition, consistent with the previous estimate. Additionally, we confirm the power-law behavior of the order parameters $\hat O \sim |J_2-J_{2c}|^{\beta}$ and correlation length $\xi \sim |J_2-J_{2c}|^{-\nu}$ close to the transition and compute the respective critical exponents $\beta$ and $\nu$, as summarized in Fig.~\ref{fig:DQCP_dynamical_overview}e and \ref{fig:DQCP_dynamical_overview}f.
We find that the average phonon displacement $u=1/N\sum_j (-1)^j u_j$ follows the same power law, indicating that the lattice displacement couples linearly to the VBS order parameter.
The order parameter critical exponent $\beta$ matches for both order parameters as required by the critical theory, and signals that the emergent U(1) symmetry between N\'eel and VBS order parameter components is preserved in the presence of spin-phonon coupling, when $\omega_0=0.6J_1$.
We find for the Luttinger parameter $K'=-\beta/\nu \approx 0.242$, in agreement with the previous estimates $K'_r$ and $K'_c$ obtained from the order parameter residuals and correlations at the critical point. 
Measurements of $\beta$ and $\nu$ as a function of frequency are shown in Appendix \ref{A:data}, Fig.~\ref{fig:DQCP_beta_nu_log_derivative_vs_K}.

Further compelling evidence for a continuous transition is found from the scaling of entanglement entropy with correlation length $\xi(\chi)$, which is expected to follow $S \sim c/6 \log(\xi)$ \cite{Pollmann2009}. Numerical data shown in Fig.~\ref{fig:DQCP_dynamical_overview}d indicate $c=1$, in agreement with a free Gaussian fixed point of the critical theory outlined before.

\begin{figure}
    \centering
    \includegraphics{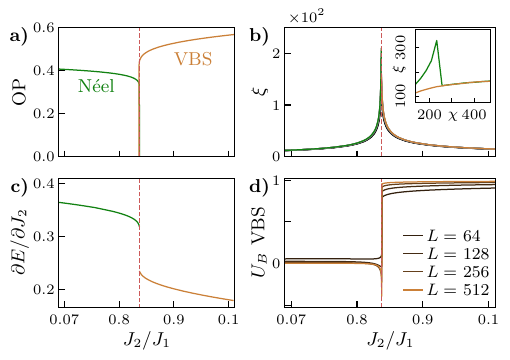}
    \caption{\textbf{Strong first-order transition in the spin-phonon model:} Below a critical phonon frequency $\omega_c$, the transition between N\'eel and VBS phase turns strongly first order. a) This is indicated by a clear jump in the order parameters, b) rapid saturation of the correlation length, and c) a discontinuity in the energy derivative at the transition. d) Further evidence for a strong-first order transition is obtained from the VBS Binder Cumulant $U_B$, which develops a diverging negative cusp with increasing system size. This behavior arises from a multi-peak distribution of the order parameter in the presence of phase coexistence. Data obtained for $\Delta_1=\Delta_2=4, g=0.25J_1, \omega_0 = 0.4J_1$.}
    \label{fig:1st_order_dynamical_overview}
\end{figure}

Below a critical frequency $\omega_0 \leq \omega_c\approx 0.45J_1$ the transition turns strongly first-order. We show a typical 1st-order cut for $\omega_0=0.4J_1$ in Fig.~\ref{fig:1st_order_dynamical_overview}. In this regime, the order parameters display a clear jump at the transition. Moreover, the MPS correlation length rapidly saturates to a finite value and the ground state energy shows a sizable discontinuity at the transition indicated by a jump in the derivative $\partial E/\partial J_2$.

To clearly distinguish the first and second order transition, we resort to an analysis of the VBS Binder cumulant \cite{Vollmayr1993} $U_B \equiv 1/2\cdot (3-\langle  \Psi^4 \rangle / \langle \Psi^2 \rangle^2)$, computed for various window sizes $L$ on the iMPS ground state. Within the ordered (disordered) phase the Binder cumulant rapidly approaches $U_B \rightarrow 1 (0)$ respectively, and $U_B$ is generally expected to be positive and scale invariant at a second-order transition. In contrast, at a first-order transition, the Binder cumulant develops a diverging negative cusp with increasing system size \cite{Vollmayr1993}. This divergence is a direct consequence of a multi-peak order parameter distribution arising from phase coexistence, and thus serves as an indicator for a (strong) 1st-order transition. We show the negative peak of the Binder cumulant of the VBS order parameter along the first-order cut in Fig.~\ref{fig:1st_order_dynamical_overview}d. In our simulations, we further define the critical frequency $\omega_c$ as the lowest frequency, for which the Binder cumulant remains monotonous. 

The numerical analysis of the phase diagram in the $\omega_0/J_1$ and $J_2/J_1$ plane for $\Delta_1=\Delta_2=4, g=0.25J_1$ is summarized in Fig.~\ref{fig:phase_diagram}, yielding the stability of the DQCP down to $\omega_0\approx 0.45J_1$. In Appendix \ref{A:data}, Fig.~\ref{fig:phase_diagram_control_cut}, we evaluate the phase diagram for an additional parameter cut $\Delta_1=4, \Delta_2=1, g=0.15J_1$, chosen such that the Luttinger parameter is $K_s\approx 0.2$ for the DQCP of the decoupled spin model. Also for this case, we find that the continuous transition breaks down at $K'\approx1/8$, and is hence independent of $K_s$.

\subsection{4-State Potts critical endpoint \& first-order regime}
\label{sec:4state-potts}

We will now describe the universality of the critical endpoint and discuss the first-order regime in more detail.
We propose that the phase diagram is described by the double sine-Gordon model, with an action given by
\begin{eqnarray}
    \begin{aligned}
        S_{\mathrm{DSG}} = \int \dd{x}\dd{\tau} \Bigg[&\frac{1}{2\pi K}\left((\partial_x\phi)^2+ (\partial_\tau\phi)^2\right) \\ 
    &- \mu\cos(4\phi) - \rho\cos(8\phi)\Bigg].        
    \end{aligned}
    \label{eq:DSG}
\end{eqnarray}
The new interaction term (last term) has the scaling dimension $[\rho] = 16K$ and is therefore relevant when $K<1/8$. The coupling is symmetry allowed, preserving the $\mathbb{Z}_2 \times U(1)$ spin-rotation and translation symmetry $\mathbf{T}_x$ and is generated by higher-order terms in the boson-fermion operator expansion as shown in Appendix \ref{a:boso}. In the regime $1/8< K <1/2$, this coupling is irrelevant and the critical line $\mu=0$ remains gapless. This justifies neglecting this contribution in the standard bosonization approaches discussed so far.
However, the presence and value of this coupling will be important for understanding the phase diagram for $\omega\leq\omega_c$.
This same effective theory has been understood to describe the universal physics and phase diagram of the Ashkin-Teller model \cite{kadanoff1980,Mussardo2004,Takacs2006,Mouland2025}. In that model the critical line is a $c=1$ CFT, which bifurcates into two Ising CFT transitions with $c=1/2$ at the 4-state Potts point.
Bifurcation occurs at the point $K=1/8$ (in our conventions) and is likewise a CFT with $c=1$.

\begin{figure}
    \centering
    \includegraphics{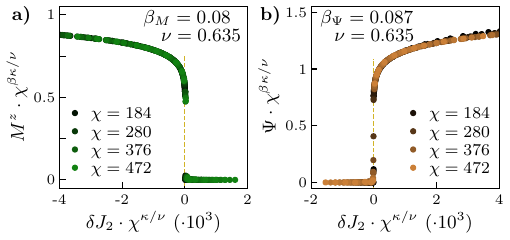}
    \caption{\textbf{Critical exponents of the transition at the endpoint $\mathbf{\omega_c}$:} We employ a finite-size scaling Ansatz to obtain  estimates for the order parameter exponent $\beta$ and correlation length exponent $\nu$ of the transition at the critical frequency $\omega_c$, where the DQCP line ends, both for the magnetization and VBS order parameter. From both we find $\beta \approx 0.0835$ and $\nu \approx 0.635$ in good agreement with the critical exponents of the four-state Potts universality class. }
    \label{fig:4potts-finite-size}
\end{figure}

First, let us use this to understand the critical endpoint of the DQCP transition line. As shown already in Fig.~\ref{fig:phase_diagram}, we measure a Luttinger parameter $K=1/8$ at this endpoint $\omega_c=0.45J_1$.
This is compatible with the picture of the continuum field theory, which tells us that the DQCP is destroyed by the presence of a second relevant coupling $\rho$ here.
Furthermore, we can identify the dynamical exponent $\eta=1/2$ via $K=1/8$ at the critical endpoint. Using a finite size scaling ansatz \cite{Tagliacozzo2008}
\begin{equation}
    \hat O \chi^{\beta \kappa / \nu} = F(\delta J_2 \chi^{\kappa/\nu})
\end{equation}
for the order parameters, we obtain precise numerical estimates for $\beta_{M}=0.08$, $\beta_\Psi=0.087$ and $\nu=0.635$, as shown in Fig.~\ref{fig:4potts-finite-size}. The numerical values are in good agreement with the exact values $\beta=1/12 \approx 0.083$ and $\nu=2/3\approx 0.667$ of the four-state Potts universality class \cite{Enting1975,Pearson1980,Nienhuis1984,Den1979}. Minor deviations are consistent with logarithmic corrections, which are expected to be small for the system sizes considered. We provide a justification of the finite size scaling ansatz in Appendix \ref{A:data} and confirm the critical exponents for a sweep along the frequency direction in Fig.~\ref{fig:4potts_finite_size_omega_direction}. 

At low phonon frequencies, the N\'eel to VBS transition is strongly first order and arises from a direct level crossing of the respective ground states. At the transition point, the two N\'eel and two VBS states become degenerate. Notably, in this regime, the VBS ground states reside on a distorted lattice ($\langle u_j \rangle \neq 0$), even at criticality. Upon increasing the phonon frequency, the lattice distortion associated with the VBS state at the transition is suppressed. The four-state Potts endpoint is reached when the distortion approaches zero and the four degenerate ground states collapse into a single symmetric critical ground state, as summarized in Fig.~\ref{fig:phase_diagram}a.

\begin{figure}
    \centering
    \includegraphics[width=0.8\columnwidth]{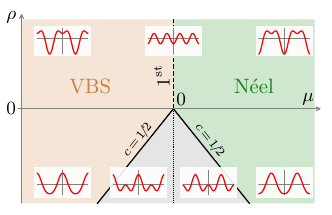}
    \caption{\textbf{Classical phase diagram of the DSG potential:} We consider a double-frequency sine-Gordon model with relevant $-\rho\cos(8\phi)$ term. For positive values of $\rho$, the classical theory implies a first-order transition between the two ordered phases, identified with the VBS and AFM phase in our setting. Instead, for $\rho<0$, an intermediate region around the original transition $\mu=0$ is found, where the minima of the potential lie in $(0,\pi/4)$, suggesting the existence of an intermediate phase. This phase is bounded by two Ising transitions.}
    \label{fig:AT-sketch}
\end{figure}

To obtain further insight into the structure of the phase transition, let us now analyze the  effective theory Eq.~\eqref{eq:DSG} at a classical level \cite{delfino2004}.
The potential term in the double sine-Gordon model
\begin{equation}
    V(\phi) = -\mu\cos(4\phi)-\rho\cos(8\phi) 
    \label{eq:dsg-with-cos-8phi}
\end{equation}
has the following minima
\begin{eqnarray}
    \begin{cases}
        \phi_0 = 0,\pi/2 & \mu>0,\rho>-\mu/4\\
        \phi_0 = \pi/4,3\pi/4 & \mu<0,\rho>-\mu/4\\
        \phi_0 = 0,\pi/4,\pi/2,3\pi/4 & \mu=0,\rho>0\\
        \phi_0 = \pm \arccos(-a/4b)/4 & \mu\neq0,\rho<-\mu/4\\
        \phi_0 = \pi/8,3\pi/8,5\pi/8,7\pi/8 & \mu=0,\rho<0\ ,\\
    \end{cases}
\end{eqnarray}
as shown in Fig.~\ref{fig:4potts-finite-size}.
Along the line $\rho = 0$, the boundary between VBS ans N\'eel order lies at $\mu=0$ and is the 1D DQCP with $c=1$ introduced in Section~\ref{sec:DQCP}.
For negative $\rho<0$, this point splits into two transitions $\pm \mu_c$ which have $\mathbb{Z}_2$ symmetry breaking and where the minimum lies in between $0$ and $\pi/4$. The transition lines flow to infrared Ising fixed points with $c=1/2$ \cite{delfino2004}.
Conversely, when $\rho>0$, a classical analysis of the potential predict a first-order transition between AFM and VBS ground states due to the potential barrier induced by the $\cos(8\phi)$ term. These results are summarized in Fig.~\ref{fig:AT-sketch}.
 
The numerical results indicate that the sign of $\rho$ is positive, given a strongly first-order transition is seen. We further argue in Appendix \ref{a:boso} that the positive sign of $\rho$ is expected within a full non-linear bosonization treatment of the spin-phonon system: In the regime $K<1/8$, the double-frequency $\cos(8\phi)$ term is dominated by contributions arising from two-fermion terms. All relevant 2-fermion contributions carry an overall minus sign, i.e., fix $\rho>0$, which provides a microscopic justification for the observed instability of the continuous transition. 

\section{Dynamical phonons: Spectral response}
\label{sec:phonon-response}

We have numerically established the role of spin-phonon coupling on the phase diagram of the spin model. Now we turn our attention to the other side of the coin. Namely, we investigate how spin-phonon coupling affects the spectrum of phonon excitations.
In the above discussion, we used an effective theory for the spin model Eq.~\eqref{eq:SG_new} where fluctuations of the lattice modes have been integrated out. The effective model was derived in Appendix~\ref{A:SG_phonons}.
When we want to instead understand the properties of the phonons, we cannot take this simplifying step, but must instead study the full spin-phonon problem.
Theoretically, one has to use approximate perturbative approaches to understand the effect of spin-fluctuations on the phonons.
One standard method is to integrate out the spin-fluctuations, in order to derive the correction to the phonon propagator $G(q,\omega)$ \cite{luther1974,lee2019,seifert2024}.
This approach works well when one understands the behavior of the separate Hamiltonians $H_s$ and $H_p$, and the spin-phonon coupling is small such that perturbation theory is controlled.
The spin-model Eq.~\eqref{eq:HPhonons} has a simple phonon Green's function, given as a momentum-independent pole $G_0(q,\omega)=1/(\omega_0^2-\omega^2)$.
Introducing a spin-phonon coupling $g$, this is corrected at leading order to
\begin{equation}\label{eq:renormphononpert}
    G(q,\omega)=\frac{1}{\omega_0^2-\omega^2 - (g^2/m)\chi_{D}(q,\omega)}\, ,
\end{equation}
where $\chi_D(q,\omega)$ is the dimer-dimer correlation function of the spin-Hamiltonian $H_s$.
Calculating these correlation functions using effective field theory \cite{luther1974} or scaling arguments \cite{seifert2024}
allows one to connect this result to the spin-Peierls instability; in this picture it is represented by the pole of Eq.~\eqref{eq:renormphononpert} renormalizing to zero energy \cite{willsher2025}.
Additionally, the spin fluctuations cause a momentum-dependent self energy to the phonon, which can be measured experimentally as a temperature-dependent Kohn anomaly.
\begin{figure}
    \centering
    \includegraphics[width=\columnwidth, trim=2 2 2 2]{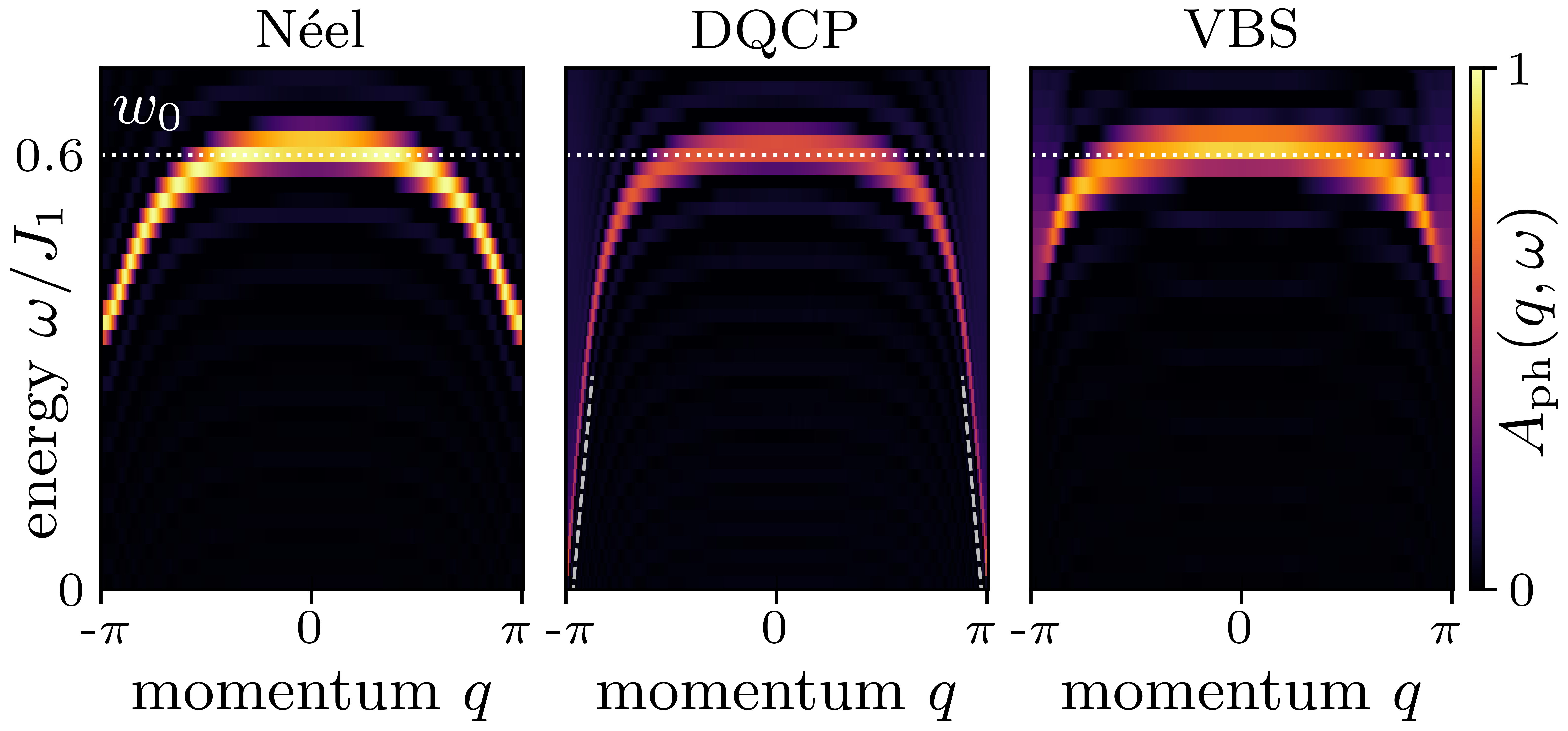}
    \caption{\textbf{Dynamical phonon response:} Single-particle phonon spectral function $A(q,\omega)$ for parameters inside the AFM phase (left), at the DQCP (middle), and inside the VBS phase (right). The white line indicates the flat phonon band of the uncoupled model. Inside both ordered phases, the spectrum remains gapped with the flat band softening at $q=\pi$, corresponding to the ordering wavevector of the spontaneous lattice staggering. At the transition, the phonon mode softens, showing linear dispersion around $q=\pi$.}
    \label{fig:spectra_overview}
\end{figure}

\begin{figure}
    \centering
    \includegraphics{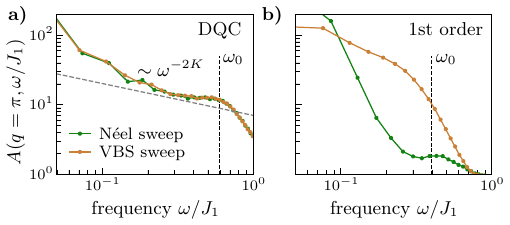}
    \caption{\textbf{Spectral weight at the transition:} a) At the DQCP, the phonon spectral function $A(k=\pi,\omega)$ at the ordering wavevector develops a low-frequency continuum with power-law scaling $A(q=\pi,\omega)\sim\omega^{-2K}$, consistent with analytical predictions. b) Conversely, at the first-order transition, we observe a large buildup of spectral weight which triggers the lattice instability. The corresponding phonon frequencies $\omega_0$ are marked by black dashed lines.}
    \label{fig:spectra_scaling}
\end{figure}

Following these general considerations, we return to our numerical analysis, where we will probe these effects in the full spin-phonon model.
To this end, we introduce the single-particle phonon spectral function
\begin{equation}
    A(q,\omega) = -\frac{1}{\pi} \text{Im}\int drdt e^{i\omega t - iqr} G(r,t),
\end{equation}
obtained as a Fourier transform of the real-space retarded Green's function
\begin{equation}
    G(r,t) = -i\pi\theta(t)\bra{\psi_0}[a(r,t),a^\dagger(0,0)]\ket{\psi_0}.
\end{equation}
We then compute $G(r,t)$ numerically by evolving the states $a^{(\dagger)}(0,0)\ket{\psi_0}$ in real time using the time dependent variational principle \cite{Haegeman2011,Haegeman2016}. We summarize details of the numerical implementation in Appendix \ref{A:numerics}. 

In Fig.~\ref{fig:spectra_overview} we show the phonon spectral function at various $J_2/J_1$ for $\omega_0=0.6J_1$ above the critical frequency $\omega_{c}$.
We take the values $J_2/J_1\in\{0.157,0.25733,0.357\}$, corresponding to a ground state in the AFM phase, at the DQCP, and in the VBS phase. Inside the ordered phases, the spectral function remains gapped and the phonon band starts to soften around the ordering wavevector $q=\pi$.
We interpret this as a Kohn anomaly of the phonon dispersion due to large fluctuations of VBS order in the spin chain.
We symmetrize the spectrum in the VBS phase to reduce numerical artifacts due to the two-site unit cell, see details in Appendix \ref{A:numerics}. 

At the transition, the phonons become gapless and a soft mode emerges at $q=\pi$, where the spectral weight of this mode is spread out into a weak continuum at low energies.
We plot the frequency dependence of this continuum on a logarithmic axes $A(q=\pi,\omega)$ and compare it to a power-law in  Fig.~\ref{fig:spectra_scaling}a.
For frequencies $\omega<\omega_0$, the numerical data is in good agreement with the critical scaling $A(q=\pi,\omega)\sim \omega^{-2K}$. 
This is consistent with the low-energy scaling of dimer-dimer correlations, $\chi_D(q=\pi,\omega)\sim \omega^{-2K}$, and agrees with the perturbative correction to the phonon Green's function in Eq.~\eqref{eq:renormphononpert}.
The low-frequency behaviour of the phonon spectral function shows a complete breakdown of the quasiparticle picture around this momentum $q=\pi$.
Instead, the phonons hybridize with the critical VBS fluctuations at the DQCP, showing signs of fractionalization in their dynamical response \cite{lee2019,seifert2024}.
Deviations from the power-law behavior appear at very small frequencies, which we attribute to the finite time cutoff imposed by the maximal simulation time. 

For comparison, we also show the corresponding spectral weight at the first-order transition for $\omega_0=0.4J_1$ in in Fig.~\ref{fig:spectra_scaling}b.
There, we observe a significant buildup of spectral weight at low frequencies, which triggers the lattice instability.
This is consistent with the perturbative description of a spin-Peierls instability being due to phonon condensation \cite{luther1974} and the understanding of the first-order transition as a spin-Peierls instability of the DQCP \cite{Hofmeier2024}.
Additionally, there is a significant difference in weight between preparing the ground state with a sweep from the AFM and VBS phase, respectively, which we attribute to reaching distinct coexistence regimes upon approaching the transition.

\section{Discussion \& Outlook}
\label{sec:discussion}

In this work, we have determined the phase diagram of a prototypical example of 1D DQC coupled to lattice vibrations. We numerically confirm the stability of DQC for phonon frequencies above a critical threshold, as predicted by Ref.~\cite{Hofmeier2024}. We demonstrate that the main consequence of spin-lattice coupling is a reduction of the Luttinger parameter $K$. The breakdown of DQC is then caused by an additional symmetry allowed $\cos(8\phi)$ term arising from higher-order bosonization.

The phenomenology of the transition below the critical phonon frequency depends on the sign of the $\cos(8\phi)$ term. Non-linear bosonization of the coupled spin-phonon model predicts a negative prefactor, which introduces a potential barrier at the transition and results in a strong first-order transition in agreement with numerical results. Conversely, with a positive prefactor, the theory is expected to describe the critical line of the Ashkin-Teller model consisting of two Ising transitions instead. The analogy to the high-symmetry critical point of the Ashkin-Teller model allows us to predict that the critical endpoint lies in the four-state Potts universality class. 

Our results are directly applicable to a broad class of 1D systems hosting DQC coupled to appropriate lattice degrees of freedom, as many such models are described by the same effective low-energy theory.
Notably, 1D systems hosting DQC offer a natural platform \cite{Lee2023} to realize and probe Ashkin-Teller type physics along the line of continuous DQC. An interesting question is whether spin-lattice coupling also affects DQCPs that do not involve translation symmetry breaking, given that the dominant effect is a renormalization of the \emph{short-range} action and an associated reduction of the Luttinger parameter. One example is the direct x-FM to z-FM transition in Ref.~\cite{Roberts2019}, where the effective theory is of BKT type and where even the existence of a dimerized phase in the adiabatic limit is elusive. Long-range 1D spin chains may provide another possible setting, as they enable breaking multiple distinct spin symmetries such as the U(1) spin-rotation symmetry \cite{Maghrebi2017}, albeit unclear if a transition between such ordered phases would be first order in the conventional Landau-Ginzburg sense \cite{Sandvik2010}.

Returning to the original 2D setting in which DQCP was introduced \cite{Senthil2004, Senthil2004prb}, we highlight that there has been an accumulation of evidence over the last years that the critical point is a non-unitary CFT. In a lattice model, this means that the critical point (described by a non-compact CP1 gauge theory) is weakly first order, with pseudo-critical RG flow \cite{Wang2017} and an approximate emergent symmetry \cite{serna2019}.
Recent fuzzy sphere studies \cite{zhou2024} onto the dual SO(6) non-linear sigma model with Wess--Zumino--Witten topological term provide evidence for this.
A particularly clear understanding of this phenomenon can be gained by extending a $d=2+\epsilon$ expansion of the SU(2) symmetric 1D model, where fixed point annihilation leads to the complex CFT \cite{nahum2020,ma2020}.
In the context of lattice theories, the continuous transition could then only be realised at a multicritical point \cite{yang2022,takahashi2024,li2026a}, which sits at the end of an extended spin liquid phase;n see Ref.~\cite{senthil2023a} for a review.
Beyond that, previous works have discussed  how spin-phonon coupling leads to a strong first order transition, using both field-theoretic \cite{Hofmeier2024} and numerical methods \cite{gotz2024tuning}.
These works and the present manuscript further complement recent studies predicting analogous lattice instabilities in 2D spin liquids \cite{seifert2024,Ferrari2025}.
The advantage of the 1D spin model studied here is that the nature of the underlying DQCP is well described by standard bosonization techniques. 
We are then able to extend these to fully understand the nature of the phase diagram and universality class of the critical endpoint.

A further advantage of the 1D setting is that quasi-exact numerical results $A(q=\pi,\omega)\sim\omega^{-2K}$ can be obtained for the critical scaling of the phonon spectral function along the DQC line. Similar scaling behavior has recently been explored in the context of spin liquids \cite{willsher2025a}, and is expected to extend to the 2D case of DQC as well. In the low-temperature regime $\beta \gg \omega_0$, the phonon spectral function is directly related to the structure factor $S(q,\omega) \approx A(q,\omega)$ via the fluctuation-dissipation theorem \cite{Mahan2013}. The latter is experimentally accessible via inelastic neutron scattering \cite{Squires1996,Shirane2002,Willis2017}, providing a direct experimentally accessible route to investigate lattice driven breakdown of DQC. In this context, it is worth noting that experiments on SrCu\textsubscript{2}(BO\textsubscript{3})\textsubscript{2} have reported a direct AFM-VBS transition whose character changes with pressure \cite{Cui2023,Guo2025}. 
Our results highlight that further theoretical work is also needed into understanding the phase diagrams of 2D realizations of (proximate) DQC with spin-phonon coupling, given the effect of pressure in modifying phonon energy scales.

On a technical level, we have been able to derive the sign of the $\cos(8\phi)$ term from microscopics, by performing a full non-linear bosonization of the coupled spin-phonon system. Our scaling arguments indicate that this term is dominated by 2-fermion contributions in the regime $K<1/8$ of relevance. This raises the question to what extent the sign of this term can be tuned and whether an intermediate coexistence phase analog to the conventional Ashkin-Teller model can emerge beyond the critical endpoint of a line of DQC. As a result, DQC would not be destroyed by the usual first order transition but rather by turning into a split transition with an intermediate phase. Amusingly, the resulting phase diagram would resemble the generic Ginzburg-Landau behavior albeit with a very different origin. Indeed, Ref.~\cite{Roberts2019} reports a splitting of the DQC line into two Ising-transitions enclosing a narrow intermediate phase. Their model realizes a DQCP between a z-ferromagnet and valence-bond solid and one may suspect the origin of this intermediate phase in the ferromagnetic couplings, given that a ferromagnetic density term generates a positive prefactor of the $\cos(8\phi)$ term. Naively, one might then expect that further introducing spin-lattice coupling introduces additional terms with a \textit{negative} prefactor, again rendering the instability 1st order. As a consequence, the intermediate phase would survive only in a narrow region around the endpoint. Understanding the interplay of such an unusual instability of DQC beyond the one-dimensional settings will be a formidable task for future numerics and analytics.

\vspace{0.5cm}
\begin{acknowledgments}
We gratefully acknowledge discussions with Johannes Hauschild, Lukas Janssen, Urban Seifert, and Manuel Weber. We acknowledge support from the Deutsche Forschungsgemeinschaft (DFG, German Research Foundation) under Germany’s Excellence Strategy–EXC–2111–390814868, TRR 360 – 492547816 and DFG grants No. KN1254/1-2, KN1254/2-1, the European Union (grant agreement No 101169765), as well as the Munich Quantum Valley, which is supported by the Bavarian state government with funds from the Hightech Agenda Bayern Plus.
\end{acknowledgments}

\section*{Data availability}
Numerical data and codes that support the findings of this article are available on Zenodo and from the authors upon reasonable request \cite{zenodo}.

\clearpage

\appendix

\section{Higher order bosonization of the spin-phonon model.}
\label{a:boso}

In this appendix, we first recapitulate the standard bosonization procedure for the sine-Gordon action in Eq.~\eqref{eq:SG} as found e.g. in \cite{Giamarchi2003}. We then extend the bosonization scheme to also include higher orders in the boson-fermion operator identity to obtain the double frequency sine-Gordon action from Eq.~\eqref{eq:DSG} and show that the sign of the $\cos(8\phi)$ coupling is fixed to be negative in the first-order regime $K<1/8$. To derive this result, we proceed in three steps: I) We recapitulate the general form of the left and right moving fermion operators $\psi_{L(R)}$ and derive the general form of two- and four-fermion operators in terms of $\psi_{L(R)}$. II) We derive the exact form of all two-fermion terms and show that the sign of $\cos(8\phi)$ is fixed to be negative for all terms. III) We introduce a general scaling argument to show that the four-fermion terms constitute \textit{subleading} contributions in the relevant regime $K<1/8$.
This result provides the theoretical basis for the arguments used in \ref{sec:4state-potts}. Since the coupling is fixed to the opposite sign as found in the Ashkin-Teller model, we do not observe the splitting into two Ising lines. Instead, these lines collapse onto a single first-order transition.

\subsection{Linear bosonization}
We start by sketching the usual steps to bosonize the spin part of  the Hamiltonian \eqref{eq:H-total}. 
First, the spin operators are fermionized using a Jordan-Wigner transformation
\begin{align}
    \begin{split}
        S_j^z &= c_j^\dagger c_j-\frac{1}{2} \,,  \\
        S_j^+ &= c_j^\dagger \exp\left(i\pi\sum_{k<j} c_k^\dagger c_k\right)\,, \\
        S_j^- &= c_j \exp\left(i\pi\sum_{k<j} c_k^\dagger c_k\right)\,,
    \end{split}
\end{align}
with a subsequent transformation $c_j \rightarrow (-1)^jc_j$ to bring the antiferromagnetic spin Hamiltonian into the fermion Hamiltonian
\begin{eqnarray}
    \begin{aligned}
        H_s = &-\frac{J_1}{2}\sum_j \bigg[(c_{j+1}^\dagger c_j + c_j^\dagger c_{j+1}) \\
        &- 2\Delta (c_j^\dagger c_j - \frac{1}{2})(c_{j+1}^\dagger c_{j+1} -\frac{1}{2})\bigg] \\
        +\frac{J_2}{2}\sum_j &\bigg[(c_{j+2}^\dagger c_j + c_j^\dagger c_{j+2})(1-2c_{j+1}^\dagger c_{j+1}) \\
        &+ 2\Delta(c_{j+2}^\dagger c_{j+2}-\frac{1}{2})(c_j^\dagger c_j - \frac{1}{2})\bigg]\,.
    \end{aligned}
    \label{eq:boso-spin-H}
\end{eqnarray}
Next, the continuum limit
\begin{align}
    \sum_j \longrightarrow \frac{1}{a}\int dx \,,\qquad c_j\longrightarrow \sqrt{a} \psi(x_j)
\end{align}
is taken and the resulting Hamiltonian is bosonized by splitting the oscillating fermionic field into left and right moving fermions according to
\begin{eqnarray}
    \begin{aligned}
    \psi(x) &= e^{ik_Fx}\psi_R(x) + e^{-ik_Fx}\psi_L(x) \\
    \psi_{L[R]} &= \frac{1}{\sqrt{2\pi\gamma}}e^{-[+]i\phi(x)-i\theta(x)}        
    \end{aligned}
    \label{eq:boso-psiLR-linear}
\end{eqnarray}
that admit a low-energy expansion into bosonic modes. Here, we introduced the UV lattice cutoff $\gamma$ of the low-energy expansion. For the antiferromagnetic spin model, the fermions are at half filling $k_F=\pi/(2a)$. One subsequently inserts Eq.~\eqref{eq:boso-psiLR-linear} into Eq.~\eqref{eq:boso-spin-H} and keeps only non oscillating terms to bring the Hamiltonian into the sine-Gordon form in Eq.~\eqref{eq:SG}. The four-fermion terms contribute to the $\cos(4\phi)$ term. Specifically, 
\begin{eqnarray}
    \begin{aligned}
        J_1\Delta c_j^\dagger c_j c_{j+1}^\dagger c_{j+1} &\rightarrow -2J_1\Delta \cos(4\phi)\\
        -J_2(c_{j+2}^\dagger c_j + c_j^\dagger c_{j+2})(c_{j+1}^\dagger c_{j+1}) &\rightarrow 4J_2\cos(4\phi)\\
        J_2\Delta c_{j+2}^\dagger c_{j+2}c_j^\dagger c_j &\rightarrow 2J_2\Delta\cos(4\phi)\,,
    \end{aligned}
\end{eqnarray}
from which $\mu = 2J_1\Delta-2J_2(2+\Delta)$ is recovered up to prefactors including the lattice spacing $a$ and UV lattice cutoff $\gamma$, see e.g. \cite{Mudry2019} for the detailed derivation. Tuning $J_2$ can induce a sign change of $\mu$ which describes the DQCP of the spin model.

\subsection{Non-linear bosonization}
We now bosonize the fermionic Hamiltonian, keeping track of higher order contributions. The boson-fermion operator to all orders reads \cite{Sachdev2011}
\begin{equation}
    \psi(x) = \sum_{m\text{ odd}}A_m e^{imk_Fx+im\phi(x)-i\theta(x)}
    \label{eq:boso-Psi}
\end{equation}
where $m$ runs over all positive and negative odd numbers and $A_m>0$ are some microscopic constants which one may calculate exactly by evaluating and comparing bosonic and fermionic correlation functions. We now explain the splitting into right- and left-movers in the higher-order theory.

In the case of a spin chain, the fermions are at exactly half-filling, i.e. we have $k_F = \frac{\pi}{2a}$, $x=ja$ and we can write \eqref{eq:boso-Psi} as
\begin{eqnarray}
    \psi(x) &= \sum_{m \in \mathbb{Z}}e^{i(2m+1)k_Fx}A_{2m+1}e^{i(2m+1)\phi(x)-i\theta(x)} \hspace{6mm}
\end{eqnarray}

\newpage

Using that $e^{2imk_Fx}=(\pm1)^x$ for $m$ even (odd) for the spin-chain which has fermions at half filling, we now split the sum according to
\begin{align}
    \begin{split}
    \psi(x) &= e^{ik_Fx}\sum_{m\text{ even}} A_{2m+1}e^{i(2m+1)\phi(x)-i\theta(x)} \\
    & + e^{-ik_Fx}\sum_{m\text{ odd}}A_{2m+1}e^{i(2m+1)\phi(x)-i\theta(x)}\,,
    \end{split}
\end{align}
and make the identifications
\begin{align}
    \begin{split}
        \Psi_R(x) &= e^{-i\theta(x)}\sum_{m\text{ even}} A_{2m+1}e^{i(2m+1)\phi(x)}\,, \\
        \Psi_L(x) &= e^{-i\theta(x)}\sum_{m\text{ odd}}A_{2m+1}e^{i(2m+1)\phi(x)}\,.
    \end{split}
    \label{eq:boso-psiLR-full}
\end{align}
We use capital symbols $\Psi_{L[R]}(x)$ to emphasize that non-linear orders are considered in the approximation of $\psi(x)$ compared to \eqref{eq:boso-psiLR-linear}. The usual expressions for $\psi_{L[R]}(x)$ are recovered by keeping only the $m=0$ term for the right mover and the $m=-1$ term for the left-mover.

\begin{widetext}

\subsubsection{General form of 2- and 4-fermion terms}

We now bosonize \eqref{eq:boso-spin-H} using \eqref{eq:boso-psiLR-full}. This creates terms that involve fields at different positions. Therefore, we must take into account the non-trivial commutator
\begin{equation}
    \comm{\phi(x_1)}{\theta(x_2)} = \frac{i\pi}{2}\text{sgn}(x_1-x_2)\,.
    \label{eq:boso-commutator}
\end{equation}
which modifies the product of exponentials to $e^{A}e^{B} = e^{A+B}e^{\comm{A}{B}/2}$. Here, we define $\text{sgn}(0)=0$. To be more explicit, we consider the following product
\begin{align}
    \begin{split}
    \Psi_{r_1}^\dagger(x_1)\Psi_{r_2}(x_2) &= \sum_{m_{r_1},n_{r_2}} A_{2m+1}A_{2n+1} e^{-i(2m+1)\phi(x_1)+i\theta(x_1)}e^{i(2n+1)\phi(x_2)-i\theta(x_2)} \\
    &= \sum_{m_{r_1},n_{r_2}} A_{2m+1}A_{2n+1}  e^{-i(2m+1)\phi(x_1)+i(2n+1)\phi(x_2)+i\theta(x_1)-i\theta(x_2)}e^{-\frac{i\pi}{2}(m+n+1)\text{sgn}(x_1-x_2)}
    \end{split}
    \label{eq:boso-2fermion-general}
\end{align}
where we use the notation $r_i = R\,(L)$ and $m_{r_1},n_{r_2}$ denotes whether the sum over $m$ and $n$ runs over even (odd) integers. The last factor comes from
\begin{equation}
    [-i(2m+1)\phi(x_1)+i\theta(x_2),i(2n+1)\phi(x_2)-i\theta(x_2)] = -\pi i(m+n+1)\text{sgn}(x_1-x_2)\,.
\end{equation}
In a next step, one may notice that typically we have $x_2 = x_1 + ka$ where $k$ is some small integer. So, one may expand $\phi(x_2) = \phi(x_1) + ka\partial_x \phi(x_1)$ (analogously for $\theta$), thereby generating derivative terms for $\phi$ and $\theta$, as well as $\cos(2l\phi), \; l=1,2,...$ interactions after appropriate grouping of the exponentials $e^{i(2n-2m)\phi(x)}$. Since we are only interested in the latter, we take
\begin{align}
    \Psi_{r_1}^\dagger(x_1) \Psi_{r_2}(x_2) = \sum_{m_{r_1},n_{r_2}} A_{2m+1}A_{2n+1}  e^{-i(2m-2n)\phi(x_1)}e^{-\frac{i\pi}{2}(m+n+1)\text{sgn}(x_1-x_2)}+ \text{derivative terms}\,.
    \label{eq:boso-2fermion-short-range}
\end{align}    

We note that all relevant 2-fermion terms are of the form \eqref{eq:boso-2fermion-short-range}. Similarly, all appearing 4-fermion terms can be brought into the general form
\begin{align}
    \begin{split}
        &\Psi_{r_1}^\dagger(x_1)\Psi_{r_2}(x_2)\Psi_{r_3}^\dagger(x_3)\Psi_{r_4}(x_4) \approx \sum_{m_{r_1},n_{r_2},p_{r_3},q_{r_4}} A_{2m+1}A_{2n+1}A_{2p+1}A_{2q+1} e^{-2i(m-n+p-q)\phi(x)} \times U(m,n,p,q)\,,
    \end{split}
    \label{eq:boso-4fermion-short-range}
\end{align}
where $U(m,n,p,q)$ constitutes a phase factor originating from the non-trivial commutator in \eqref{eq:boso-commutator}.

\subsubsection{Contributions to $\cos(4\phi)$ and  $\cos(8\phi)$: 2-fermion terms}

We now compute the contributions to the prefactors $\mu$ and $\rho$ arising from the 2-fermion terms of the spin Hamiltonian in \eqref{eq:boso-spin-H} in non-linear bosonization. Starting from \eqref{eq:boso-2fermion-short-range}, the sums can be simplified by introducing an auxiliary summation variable $2l = 2n-2m$. For example, in the case $r_1=r_2=R$, we obtain
\begin{eqnarray}
    \begin{aligned}
        &\Psi_R^\dagger(x_1)\Psi_R(x_2) = \sum_{l\text{ even}} e^{-i\frac{\pi}{2}(l+1)\text{sgn}(x_1-x_2)}\sum_{m\text{ even}}A_{2m+1}A_{2m+1+2l} \cos(2l\phi)\\
        &\approx i^{\text{sgn}(x_1-x_2)} \Bigg[2\left(\sum_{m\;\text{even}}A_{2m+1}A_{2m+5}\right)\cos(4\phi) + (-i)^{\text{sgn}(x_1-x_2)}\left(2\sum_{m\;\text{even}}A_{2m+1}A_{2m+9}\right)\cos(8\phi)\Bigg]\,.
    \end{aligned}
    \label{eq:boso-summation-variable}
\end{eqnarray}
Collecting the sums over $A_m$ in two constants $C_1^{RR},C_1^{LL} > 0$, we obtain
\begin{equation}
    \Psi_R^\dagger(x_1)\Psi_R(x_2) \approx \begin{cases}
        C_1^{RR}\cos(4\phi)+C_2^{RR}\cos(8\phi)  &\text{if } \;\; x_1=x_2\\
        i^{\text{sgn}(x_1-x_2)}\left[C_1^{RR}\cos(4\phi)-C_2^{RR}\cos(8\phi)\right] & \text{else}
    \end{cases}
\end{equation}
Similarly,
\begin{equation}
    \Psi_L^\dagger(x_1)\Psi_L(x_2) \approx \begin{cases}
        C_1^{LL}\cos(4\phi)+C_2^{LL}\cos(8\phi) & \text{if }\;\; x_1=x_2\\
        (-i)^{\text{sgn}(x_1-x_2)}\left[C_1^{LL}\cos(4\phi)-C_2^{LL}\cos(8\phi)\right] &\text{else}
    \end{cases}
\end{equation}
with $C_1^{LL},C_2^{LL}>0$. We also note the result for the cross terms
\begin{align}
    \begin{split}
        \Psi_R^{\dagger}(x_1)\Psi_L(x_2) &\approx -2iC_1^{RL}\sin(2\phi) +2iC_2^{RL}\sin(6\phi) \\
        \Psi_L^{\dagger}(x_1)\Psi_R(x_2) &\approx 2iC_1^{RL}\sin(2\phi) -2iC_2^{RL}\sin(6\phi)\,,
    \end{split}
    \label{eq:A4-RL-terms}
\end{align}
which contribute to the spin-phonon coupling.

Summarizing the quadratic terms, we find that the individual fermionic terms contribute as follows to the interactions:
\begin{subequations}
    \begin{align}
        \frac{J_1}{2}(c_{j+1}^\dagger c_j + c_j^\dagger c_{j+1}) &\longrightarrow J_1\left[C_1\cos(4\phi) - C_2\cos(8\phi)\right] \\
        \frac{J_2}{2}(c_{j+2}^\dagger c_j + c_j^\dagger c_{j+2}) &\longrightarrow J_2\left[C_1\cos(4\phi) - C_2\cos(8\phi)\right] \\
        -\frac{J_1\Delta}{2}(c_{j+1}^\dagger c_{j+1} + c_j^\dagger c_j) &\longrightarrow J_1\Delta \left[-C_1\cos(4\phi) +C_2\cos(8\phi)\right] \\
        -\frac{J_2\Delta}{2}(c_{j+2}^\dagger c_{j+2} + c_j^\dagger c_j) &\longrightarrow J_2\Delta \left[-C_1\cos(4\phi) + C_2\cos(8\phi)\right]
    \end{align}
\end{subequations}
where we defined new constants $C_i=C_i^{RR}+C_i^{LL}>0$.
Comparing with \eqref{eq:boso-spin-H}, we obtain
\begin{eqnarray}
    \begin{aligned}
    &-C_1J_1[1-\frac{J_2}{J_1}-\Delta(1+\frac{J_1}{J_2})]\cos(4\phi) \\
    &- C_2J_1/2 \left[\frac{J_2}{J_1}-1+\Delta(1+\frac{J_2}{J_1})\right]\cos(8\phi)\,. 
    \end{aligned}
    \label{eq:appboso_cos4-8p}
\end{eqnarray}
for the contributions to $\mu$ and $\rho$ respectively. Thus, when including higher orders, an additional $\cos(8\phi)$ term is already generated from the quadratic terms of the spin part of the Hamiltonian \eqref{eq:H-total}. The coupling strength of this term is strictly negative in the relevant regime $\Delta>1$, implying $\rho>0$ in Eq.~\ref{eq:dsg-with-cos-8phi}. We now show that this remains the case when including contributions from spin-phonon coupling.
\end{widetext}

\subsubsection{Contributions to $\cos(8\phi)$: Spin-phonon coupling}

We further have to consider leading contributions from spin-phonon coupling, given that the prefactor of these terms scales with $1/\omega_0^2 \xrightarrow[]{\omega_0\rightarrow0} \infty$ in the adiabatic limit. When bosonizing the spin-phonon interaction, we have to account for the instability towards a staggered lattice distortion. In terms of the phonon displacement field, this corresponds to the condition $u(x+a) = -u(x)$. Comparing with \eqref{eq:H-total} and \eqref{eq:action-with-phonons}, we bosonize the 2-fermion terms of the staggered component
\begin{align}
    \begin{split}
    &(-1)^x\bm{S}(x)\cdot \bm{S}(x+a) \\
    &= -i\psi_R^\dagger(x)\psi_L(x+a) + i\psi_L(x)^\dagger \psi_R(x+a) + \text{h.c.}
    \end{split}
\end{align}
We argue below that these are indeed the leading contributions to the staggered spin bilinear.
Using \eqref{eq:A4-RL-terms}, this operator can be written as
\begin{align}
    \begin{split}
        i\psi_R^\dagger(x)\psi_L(x+a) = C_1 \sin(2\phi) - C_2\sin(6\phi)
    \end{split}
\end{align}
After integrating out the phonons in \eqref{eq:action-with-phonons}, we are left with
\begin{multline}
    -g^2\int dx d\tau d\tau'G_{\omega_0}(\tau-\tau') \times \\
    \left[C_1 \sin(2\phi) + C_2\sin(6\phi)\right] \times \\
    \left[C_1 \sin(2\phi') + C_2\sin(6\phi')\right]
\end{multline}
and using the usual trigonometric identity we rewrite the resulting terms as
\begin{align*}
    \begin{split}
        \sin(2\phi)\sin(2\phi') &= \frac{1}{2}\left[\cos(2(\phi-\phi'))-\cos(2(\phi+\phi'))\right] \\
        \sin(6\phi)\sin(2\phi') &= \frac{1}{2}\left[\cos(6\phi-2\phi')-\cos(6\phi+2\phi')\right] \\
        \sin(2\phi)\sin(6\phi') &= \frac{1}{2}\left[\cos(2\phi-6\phi')-\cos(2\phi+6\phi')\right] \\
        \sin(6\phi)\sin(6\phi') &= \frac{1}{2}\left[\cos(6(\phi-\phi'))-\cos(6(\phi+\phi'))\right]\,.
    \end{split}
\end{align*}
We can readily read off that the $\cos(8\phi)$ interactions are always negative, i.e., the phonons generate higher order terms such as
\begin{equation}
    -C_1C_2\frac{g^2}{\rho\omega_0^2} \cos(8\phi)
\end{equation}
which dominate in the adiabatic limit $\omega_0\rightarrow 0$. 

\subsubsection{Leading contributions from nonlinear terms.}

We now demonstrate that contributions from 4-fermion terms in \eqref{eq:boso-4fermion-short-range} are subleading compared to contributions originating from the 2-fermion terms in \eqref{eq:boso-2fermion-short-range} in the first-order regime $K<1/8$ of interest. 

First, we note that the scaling dimension of the boson-fermion operator in \eqref{eq:boso-Psi} should remain well defined upon including higher order terms. The scaling dimensions $[e^{in\phi}] = n^2K/4$, $[e^{im\theta}] = m^2/(4K)$ imply that the operator $e^{im\phi-i\theta}$ transforms under an RG step $\gamma \rightarrow \gamma'=\gamma/b$ as $A_me^{im\phi-i\theta} \rightarrow A_m'b^\Delta e^{im\phi'-i\theta'}$, where $\Delta = m^2K/4+1/(4K)$. Keeping track of the dimensionality constant $d=1$, this implies that the prefactors scale as
\begin{equation}
    A_m \sim \gamma^{m^2K/4+1/(4K)-d}\,.
    \label{eq:boso-Am-scaling}
\end{equation}
with the UV lattice cutoff $\gamma \ll 1$. For $m\gg1/K$, the prefactors $A_m \sim \gamma^{m^2K/4}$ are dominated by $m$ and fall of exponentially, ensuring that all sums introduced above converge. 

The 4-fermion terms in \eqref{eq:boso-4fermion-short-range} can also be grouped in terms of $\cos(2l\phi),\;\sin(2l\phi), \; l\in\mathbb{N}$ by introducing an auxiliary summation variable
\begin{eqnarray}
        2n-2m+2q-2p &= 2l\,,
    \label{eq:boso-constraints-sums}
\end{eqnarray}
similar to $2n-2m=2l$ in \eqref{eq:boso-summation-variable} for the 2-fermion terms. We can then determine how the largest terms in each sum from \eqref{eq:boso-2fermion-short-range} and \eqref{eq:boso-4fermion-short-range} contributing to $\cos(2l\phi)$, given by
\begin{eqnarray}
    \begin{aligned}
    &A^{(2)}\equiv \max_m A_{2m+1}A_{2m+1+2l} \quad \text{and} \\
    &A^{(4)}\equiv \max_{m,n,p} A_{2m+1}A_{2n+1}A_{2p+1}A_{2(m-n+p)+1+2l}
    \end{aligned}
\end{eqnarray}
scale with the UV cutoff $\gamma$. From \eqref{eq:boso-constraints-sums}, we have that
\begin{widetext}
\begin{align}
    \begin{split}
        A_{2m+1}A_{2m+1+2l} &\sim \gamma^{f^{(2)}_lK/4+2/(4K)-2} \\
        A_{2m+1}A_{2n+1}A_{2p+1}A_{2(m-n+p)+1+2l} &\sim \gamma^{f^{(4)}_lK/4+4/(4K)-4}
    \end{split}
    \label{eq:largest-terms}
\end{align}
where we introduced the functions
\begin{align}
    \begin{split}
        f^{(2)}_l &\equiv (2m+1)^2+(2m+1+2l)^2 \\
        f^{(4)}_l &\equiv (2m+1)^2+(2n+1)^2+(2p+1)^2+(2m-2n+2p+1+2l)^2\,.
    \end{split}
    \label{eq:boso-fs}
\end{align}    

\end{widetext}
For fixed $K$, the minima of $f^{(2)}_l$ and $f^{(4)}_l$ correspond to the term scaling with the lowest exponent in the UV lattice cutoff $\gamma$ and thus to the largest term in the respective sum. We summarize these minima for $l=1,2,3,4$ in Table~\ref{t:exponents}.

In order to determine the leading contributions to the prefactors of the resulting $\cos(2l\phi)$ terms, we further have to account for the constant factors $2/(4K)-2$ for the 2-fermion terms and $4/(4K)-4$ for the 4-fermion terms in \eqref{eq:largest-terms}. 

For $K\rightarrow0$, the constant terms dominate, and we have $A^{(2)}\sim \gamma^{1/(2K)} \gg \gamma^{1/K}\sim A^{(4)}$, implying that 2-fermion terms are the leading contributions to the prefactors.

Conversely, for large $K$, the terms $f_l^{(*)}K/4$ dominate the exponent (Consider, for example, $K=1/4$, where $2/(4K)-2=4/(4K)-4=0$). From Table~\ref{t:exponents}, we see that $\min f^{(4)}_l< \min f_l^{(2)}$ for $l=2,3,4$, implying that the 4-fermion terms constitute the leading contributions to the relevant prefactors.

Finally, we consider $K=1/8$ which sits between these two extremes. There, we expect all terms contribute significantly. We summarize the exponents for $A^{(2)}$ and $A^{(4)}$ in Table~\ref{t:exponents}, where we see that the 2-fermion contributions are generally larger. This means in the whole region $K \leq 1/8$, the 2-fermion terms will dominate the contribution to $\rho$.

Let us summarize what we found: In the first-order regime $K\leq 1/8$ of the spin-phonon model, the 4-fermion terms constitute \textit{subleading} contributions to the overall prefactor $\rho$ of the $\cos(8\phi)$ term in \eqref{eq:DSG}. This justifies considering only 2-fermion terms to determine the overall sign of $\rho$. Conversely, in the DQCP regime $K>1/8$, contributions from 4-fermion terms generally provide an equal or dominant contribution to the overall prefactor. Note that this is an important justification for the sine-Gordon action in \eqref{eq:SG}, where tunability of $\mu$ is introduced by contributions from 4-fermion terms. (In fact, the sign of $\mu$ in \eqref{eq:appboso_cos4-8p} would be fixed when considering higher-order 2-fermion terms, but these contributions are \textit{subleading} in the DQCP regime.)

\begin{center}
    \begin{table}[h]
        \centering
        \begin{tabular}{r|c|c|c|c|}
            $\mathbf{2l}$ & $\mathbf{2}$ & $\mathbf{4}$ & $\mathbf{6}$ & $\mathbf{8}$  \\ \hline
            $\min f_l^{(2)}$ & 2 & 10 & 18 & 34 \\ \hline
            $\min f_l^{(4)}$ & 4 & 4 & 12 & 20 \\ \hline
            2-fermion exponent at $K=1/8$ & 2.06 & 2.31 & 2.56 & 3.06 \\ \hline
            4-fermion exponent at $K=1/8$ & 4.125 & 4.125 & 4.375 & 4.625 \\ \hline
        \end{tabular}
    \caption{Summary of minima for $f^{(2)}_l$ and $f^{(4)}_l$ as defined in \eqref{eq:boso-fs}. At $K=1/8$, contributions to $\cos(2l\phi)$ from various terms carry a prefactor that scales with the shown exponent in the lattice cutoff $\gamma$. 2-fermion contributions dominate. See main text for details.}
    \label{t:exponents}
    \end{table}
\end{center}

\section{Numerical implementation}
\label{A:numerics}

All simulations are performed using the TeNPy library \cite{Hauschild2024}. We explicitly enforce conservation of the total spin-z component $\sum_j S^z_j$. We implement the Hamiltonian in Eq.~\eqref{eq:J1J2XXZ_und}  with one spin-1/2 per site and the Hamiltonian of the spin-phonon model in Eq.~\eqref{eq:H-total} as a chain of alternating spin-1/2 and boson sites. For the spin model, it suffices to consider bond dimensions $\chi_s\leq220$, whereas for the spin-phonon model, we consider bond dimensions up to $\chi=472$. 

We implement a sweep protocol to capture the transition. The primary advantage is improved convergence close to the critical point. In addition, the protocol induces controlled hysteresis near criticality, which can be used to obtain a precise estimate of the critical point, as outlined below. We first initialize a trial MPS $|\psi_0 \rangle$ in the fully polarized N\'eel state or a chain of singlets on neighboring bonds and compute the ground state for some value of $J_2$ deep inside the corresponding ordered phase. We use the infinite system variant of the Density Matrix Renormalization Group (DMRG) implemented within the TeNPy library \cite{Hauschild2024}.

In each step, we approach the transition by slightly varying $J_2/J_1$ and reuse the previous variational ground state as initial state for the subsequent simulation. \emph{Crucially}, we recycle the final left and right environment of the previous simulation as initial environment for the subsequent run. These environments correspond to the left and right infinite half chain. Within the DMRG protocol, we insert an additional unit cell into these environments after each sweep. In particular, this means that the "current" Hamiltonian is only implemented on a window $L_\text{cell} \cdot N_\text{sweeps}$, with $L_\text{cell}=16$.  
To ensure that simulations are converged, we insert at least $N\approx \xi_{\max}$ sites into each environment in each step and vary $J_2/J_1$ slower, when close to the transition. The main advantage of this approach is that it gives better control over the order parameter residuals discussed in the main text.

In the spin-phonon model, local DMRG updates act on a single spin site. This leads to a reduces buildup of entanglement upon approaching the transition. To circumvent this problem, we introduce a small density mixer \cite{White2005} with initial amplitude $a_0=10^{-5}$ that exponentially decays to a final amplitude $a_f = 10^{-8}$ over $20$ steps \emph{in each simulation}. The strength of the mixer is tuned to generate sufficient entanglement, while allowing for sufficient hysteresis close to the critical point.
    
A precise estimate of the critical point $J_{2c}/J_1$ is found using a variant of the method introduced in Ref.~\cite{Roberts2019}. At finite bond dimension $\chi$, an MPS does not capture the correct second-order phase transition, but instead exhibits a weakly first-order transition at a pseudo-critical point $J_{2c}(\chi)/J_1$ that shifts with $\chi$. Due to hysteresis in sweep protocols, simulations near the transition converge to a metastable state on the opposite side of the initial phase rather than the correct ground state, resulting in a level crossing. In Ref.~\cite{Roberts2019}, the critical point is estimated from this level crossing via the zero crossing of $\Delta E \equiv E_\text{VBS}(J_2)-E_\text{AFM}(J_2)$, where $E_*$ denotes the variational energy of a sweep originating in the respective phase.

\begin{figure}
    \centering
    \includegraphics{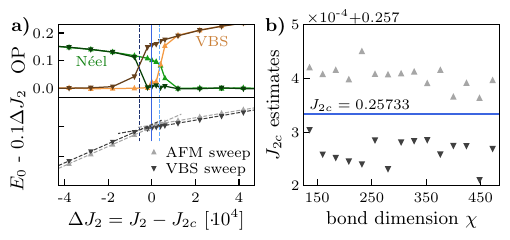}
    \caption{a) We obtain a precise estimate of the critical point $J_{2c}$ by performing separate linear fits of ground state energy $E_0$ on both sides of the transition for each sweep direction, thereby obtaining two independent estimates $J_{2c,AFM}(\chi)$ and $J_{2c,VBS}(\chi)$ for each bond dimensions $\chi$. b) The estimates $J_{2c,AFM}(\chi)$ (light gray) and $J_{2c,VBS}(\chi)$ (dark gray) show minimal drift with bond dimension, allowing us to take the arithmetic mean as final estimate for the critical point $J_{2c}$.}
    \label{fig:J2c-method}
\end{figure}

In our simulations, $\Delta E$ becomes unstable at large bond dimensions. Instead, we determine a critical point for each sweep separately by fitting the energy linearly on both sides of the transition (identified via the order parameters), with $J_{2c,\text{AFM(VBS)}}/J_1$ given by the intersection. Due to hysteresis, we find $J_{2c,AFM}(\chi)>J_{2c,VBS}(\chi)$. Both estimates show negligible drift with $\chi$ and differ by $J_{2c,AFM}(\chi)-J_{2c,VBS}(\chi) \lesssim 4\cdot 10^{-4}J_1$. We can therefore take 
\begin{equation}
J_{2c} = 1/N_\chi \sum_\chi \frac{1}{2}(J_{2c,AFM}(\chi)+J_{2c,VBS}(\chi))
\end{equation}
as our final estimate. We summarize the method in Fig.~\ref{fig:J2c-method}.

\begin{figure}
    \centering
    \includegraphics{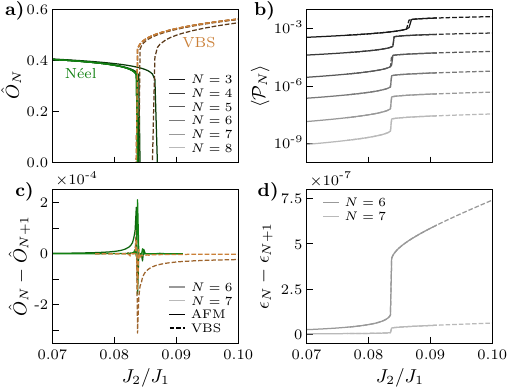}
    \caption{Effects of the phonon cutoff $N$. Results are shown for $\omega_0=0.4$, well within the 1st-order regime. a) The location of the critical point remains unchanged for $N\geq5$. b) The order parameters differ by $\hat O_N - \hat O_{N+1} \leq 3\cdot10^{-4}$, corresponding to a relative deviation $\delta \hat O /\hat O \leq 10^{-3}$. c) The ground state exhibits exponentially small overlap with configurations containing more than $N\geq3$ phonons per bond, measured via the projector $\mathcal{P}_N \equiv \ket{n_j=N}\bra{n_j=N}$. d) Near the transition, the variational ground-state energy varies only slightly with increasing cutoff $N$. Notably, the discontinuity in the ground-state energy in the first-order regime exceeds this variation by several orders of magnitude.}
    \label{fig:phonon-cutoffs}
\end{figure}

We implement dynamical phonons as bosons, Eq.~\eqref{eq:HPhonons-2nd-quant}, and truncate the local Hilbert space to at most $N_\text{max}=6$ phonons per site for computational feasibility. Here, we briefly justify this approximation. For small $\omega_0$, the dominant phonon energy scale is given by the spin-phonon coupling $a^\dagger_j+a_j$, and the phonon state is expected to be close to some coherent state $\ket{\alpha} = e^{-|\alpha|^2/2}\sum_{n=0}^\infty \alpha^n/\sqrt{n!}\ket{n}$. At any finite $\omega_0$, the average occupation $\langle n_j \rangle = \langle a^\dagger_j a_j \rangle$ remains finite and we expect the state to have maximal weight around $|\alpha|^2 \approx \langle n_j\rangle$, with exponentially suppressed tails at large enough $n\gg|\alpha|^2$. Additionally, $\langle n_j \rangle$ decreases upon increasing $\omega_0$. Numerically, we find $ \langle n_j \rangle \lesssim 0.33$ and $\langle \psi_{gs}\ket{n_j=6}\bra{n_j=6}\psi_{gs} \rangle \lesssim 5 \cdot 10^{-6}$ for frequencies $\omega_0 \geq 0.4$, hence even well inside the first-order regime, justifying the truncation. We further confirm numerically that the relevant physical behavior is unchanged even in the first order regime when tuning the phonon cutoff (where truncation is most significant), see Fig.~\ref{fig:phonon-cutoffs}.

The phonon spectral function is given by $A(q, \omega)=-1/\pi \, \text{Im}(G(q,\omega))$, where $G(q,\omega)$ denotes the Fourier transform of the retarded Green's function
\begin{equation}
    G(r,t) = -i\pi\theta(t)\bra{\psi_0}[a(r,t),a^\dagger(0,0)]\ket{\psi_0}.
\end{equation}
Here, we assume time-translation invariance and translational invariance of the ground state $\ket{\psi_0}$. In the VBS phase, the latter is broken. In principle, this requires computing the Green's function separately for each sublattice of the enlarged two-site unit cell. Owing to the substantial cost of the numerical simulations, we restrict the analysis to applying the first operator only on the first site of the unit cell and subsequently evaluate correlations with operators on both sublattices. Because the VBS ground state  breaks translation invariance only by staggering, correlations between the sublattices differ only by a factor $(-1)$. Therefore  this approximation produces the correct results, even though the spectral function inside the VBS phase is not  translation symmetric in real space.

We compute the retarded Green's function numerically using the time-dependent variational principle (TDVP) \cite{Haegeman2011,Haegeman2016}. The initial operator is applied at the center of the chain, and correlations are evaluated for all sites of the chain by applying the second operator and computing the overlap with the unevolved ground state at regular intervals $\delta t=0.2/J_1$, up to a maximal time $t_m \approx 80/J_1$.

To eliminate finite-size effects, we use a hybrid approach in which time evolution is performed on an MPS segment with $L=256$ phonon sites embedded in two infinite half chains \cite{Phien2012, Milsted2013}. The MPS segment is obtained by repeating the unit cell and the boundary environments are given by the dominant left and right eigenvector of the MPO transfer matrix up to an extensive energy contribution. These environments can be efficiently computed using the method introduced in Ref.~\cite{Phien2012}. We use the 2-site version of TDVP which allows the bond dimension to increase over time. We start from the ground state with bond dimension $\chi=400$ and limit the maximal bond dimension of the subsequent time evolution to $\chi=600$. 

We subsequently use linear prediction \cite{White2008} to extrapolate the correlations $G(r,t)$ up to twice the original simulation time. We then apply a Gaussian window function $e^{-t^2/(2\sigma^2\cdot4t_m^2)}$ with $\sigma=0.25$ to reduce Gibbs ringing. Applying this window changes the scaling of the spectral weight according to
\begin{equation}
A_\text{total}(q,\omega) = A(q,\omega) \star \hat{f}(\omega),
\end{equation}
where $\star$ denotes the convolution and $\hat{f}(\omega) \equiv \sqrt{\pi \sigma^2t_m^2}e^{-\pi^2 \omega^2\sigma^2t_m^2}$ is the Fourier transform of the Gaussian window function. The frequency broadening arising from this gaussian window is $\delta\omega = \frac{1}{\pi\sigma t_m} \approx 0.008J_1$.  

\section{Effective sine-Gordon theory for dynamical phonons}
\label{A:SG_phonons}

Here, we reproduce results from Ref.~\cite{Hofmeier2024} for the low-energy effective theory of the spin-phonon model in Eq.~\eqref{eq:H-total} in the large phonon frequency limit $\omega_0 \rightarrow \infty$. If we include temporal fluctuations of the field $u_j$, the combined action of spins and phonons reads
\begin{equation}
    S(\omega_0) = S_\text{SG} + \int dx d\tau \Big\{ \frac{\rho}{2}\big[|\partial_\tau u|^2+\omega_0^2|u|^2\big] +2gu\sin(2\phi)\Big\}
    \label{eq:action-with-phonons}
\end{equation}
with phonon mass density $\rho = m/a$ and spin-phonon coupling strength $g$. Integrating out the quadratic displacement field yields
\begin{eqnarray}
    S = S_{\mathrm{SG}} + g^2\int\dd{x}\dd{\tau}\dd{\tau'} G(\tau-\tau') \\
    \times [\cos(2\phi_{\tau}+2\phi_{\tau'})-\cos(2\phi_{\tau}-2\phi_{\tau'})].
    \label{eq:sg-phonon-integrated}
\end{eqnarray}
with the phonon propagator
\begin{equation}
    G_{\omega_0}(\tau-\tau') = \frac{1}{2m\omega_0} e^{-\omega_0\abs{\tau-\tau'}}\,.
\end{equation}
In the limit $\omega_0\to\infty$ the propagator is completely local in space and time. Thus, for large $\omega_0$, we can Taylor expand the cosine terms which depend on $\phi(\tau),\phi(\tau')$ \cite{Hofmeier2024}. The expansion of the sum reads
\begin{eqnarray}
    \cos\{2[\phi(\tau)+\phi(\tau')]\} = \cos[4\phi(\tau)] \\
    -2\Delta\tau \partial_\tau \phi\sin(4\phi(\tau)) - (\Delta\tau)^2\partial^2_\tau \phi \sin[4\phi(\tau)] \\ -2(\Delta\tau)^2(\partial_\tau \phi)^2\cos[4\phi(\tau)]
\end{eqnarray}
The term $\partial_\tau\phi\sin(4\phi)$ becomes relevant at $K<1/4$, but constitutes only a total derivative $\partial_\tau\phi\sin(4\phi) \sim \partial_\tau \cos(4\phi)$ and hence describes boundary effects which we may ignore in the thermodynamic limit. The next higher derivative terms have scaling dimension $2+4K$ and are thus irrelevant over the whole considered regime.
Inserting this back into the action Eq.~\eqref{eq:sg-phonon-integrated} renormalizes the prefactor of the $\cos (4\phi)$ term
\begin{equation}
    \mu' = \mu +\frac{g^2}{\rho\omega_0^2}
\end{equation}
implying that the DQCP simply gets shifted from $\mu=0$ to $\mu_c = -g^2/(\rho\omega_0)^2$.

In order to Taylor expand the difference, one has to normal order first. Using the identity \cite{Knops1980,nozieres1987}
\begin{equation}
    \cos(\phi) = :\cos(\phi):e^{-\langle\phi^2\rangle_0/2}
\end{equation}
and using the standard result $\langle[\phi(\tau)-\phi(\tau')]^2\rangle_0 = K\log\abs{\tau-\tau'}$ \cite{Giamarchi2003}, one has to leading order
\begin{equation}
    \cos \{2[\phi(\tau)-\phi(\tau')]\} = \Big\{ 1-2(\tau-\tau')^2[\partial_\tau\phi(\tau)]^2\Big\}\big |\tau-\tau'\big |^{-2K}
\end{equation}
which after performing the $\tau'$ integration renormalizes the Luttinger parameter $K'=\gamma K<K$
\begin{equation}
    \gamma = \Big( \frac{2\pi Kg^2 \Gamma(3-2K)}{\rho\omega_0^{4-2K}} +1 \Big)^{-1/2}    
\end{equation}
where $\Gamma(x)$ denotes the Euler gamma function. 

The new theory is of the same form as the original sine-Gordon action \eqref{eq:SG}, 
\begin{equation}
        S_{\mathrm{SG}'} = \int \dd{x}\dd{\tau} \left[\frac{1}{2\pi K'}\left(\gamma(\partial_x\phi)^2+\gamma^{-1}(\partial_\tau\phi)^2\right) 
    - \mu'\cos(4\phi)\right] \,.
    \label{eq:SG_newfull}
\end{equation}
but with renormalized parameters
\begin{align}
     K'=\gamma K \, , \quad
     \mu'=\mu+g^2/(\rho \omega_0^{\ 2})
\end{align}
and hosts a line of DQCP as long as $K'>1/8$. Solving for $\gamma K = 1/8$, we obtain
\begin{equation}
    \rho \omega_c^{4-2K} = \frac{2\pi K\,\Gamma(3-2K)}{64K^2-1}g^2
\end{equation}
This indicates that the transition is always unstable for any $g>0$ when the initial $K=1/8$, and that the critical frequency is suppressed for larger initial $K$.

\begin{figure}
    \centering
    \includegraphics{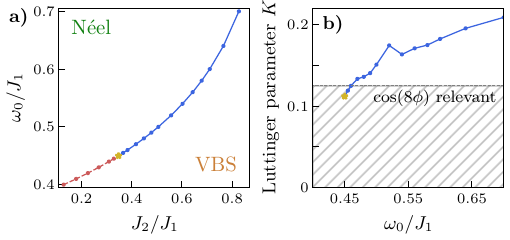}
    \caption{Phase diagram of the spin-phonon model with $\Delta_1=4, \Delta_2=1, g=0.15J_1$. In the limit $\omega_0\rightarrow\infty$, the corresponding spin model undergoes a DQCP with $K_0\approx0.20$. The line of continuous transitions terminates at $K'\approx1/8$, as in Fig.~\ref{fig:phase_diagram}, confirming that deconfined criticality breaks down at $K'=1/8$ independent of the initial Luttinger parameter $K$.}
    \label{fig:phase_diagram_control_cut}
\end{figure}


\section{Additional data}
\label{A:data}

In Fig.~\ref{fig:phase_diagram_control_cut} we present the phase diagram for an additional phase diagram with $\Delta_1=4, \Delta_2=1, g=0.15J_1$. For these parameters, the spin-only model $H_s$ undergoes a DQCP with Luttinger parameter $K\approx0.20$. Tuning $\omega_0/J_1$, we determine the critical frequency $\omega_c/J_1$ from the VBS Binder cumulant as outlined in the main text. We find that the DQCP line again terminates at $K'(\omega_c)\approx1/8$, demonstrating that the breakdown of DQC occurs at $K'=1/8$ independent of the Luttinger parameter $K$ of the underlying transition.

\begin{figure}[h]
    \centering
    \includegraphics[width=\columnwidth, trim = 0 0 0 0]{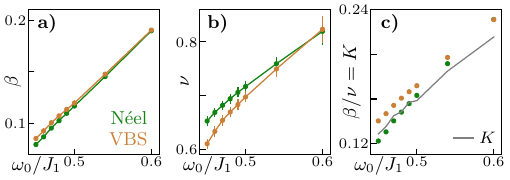}
    \caption{a,b) Critical exponents $\beta$ and $\nu$ determined via discrete log-log derivatives, for frequencies close to the critical endpoint shown in Fig.~\ref{fig:phase_diagram}. c) The estimate $K=\beta/\nu$ is consistent with previous estimates of the Luttinger parameter $K$ (gray line) obtained in the main text.}
    \label{fig:DQCP_beta_nu_log_derivative_vs_K}
\end{figure}

In Fig.~\ref{fig:DQCP_dynamical_overview} of the main text, we show that the critical order parameter exponent $\beta$ and correlation length exponent $\nu$, governing the power laws $\hat O \sim |J_2-J_{2c}|^\beta$ and $\xi \sim |J_2-J_{2c}|^{-\nu}$ are consistent with the field theory prediction $K=\beta/\nu$. In Fig.~\ref{fig:DQCP_beta_nu_log_derivative_vs_K}, we show $\beta(\omega_0)$ and $\nu(\omega_0)$ for frequencies $0.45J_1\approx\omega_c \leq \omega_0 \leq 0.6J_1$ close to the critical endpoint of the phase diagram in Fig.~\ref{fig:phase_diagram}. We determine $\beta$ and $\nu$ via the mean of the longest plateau in the discrete log-log derivative, e.g. $\beta = \partial \log (\hat O(|J_2-J_{2c}|))/\partial \log (|J_2-J_{2c}|)$. The numerical data demonstrates good agreement between the critical exponents $\beta_M$ and $\beta_\Psi$ of the order parameters along the whole frequency range. Moreover, the resulting ratios $\beta/\nu$ agree with the estimates of the Luttinger parameter $K$ obtained in the main text. 

In Sec.~\ref{sec:4state-potts}, we argue that the transition at the critical frequency $\omega_c$ lies in the 4-state Potts universality class and employ a finite-size scaling Ansatz to obtain the critical exponents $\beta$ and $\nu$. The idea behind the ansatz
\begin{equation}
    \hat O \chi^{\beta \kappa / \nu} = F(\delta J_2 \chi^{\kappa/\nu})\,,
\end{equation}
where  $\delta J_2 \equiv (J_2-J_{2c})/J_{2c}$ is as follows: At criticality, the exponent $\kappa=6/(c(\sqrt{12/c}+1))$ relates the MPS bond dimension $\chi^\kappa = \xi(\chi)$ to the MPS correlation length $\xi$ via finite entanglement scaling \cite{Tagliacozzo2008, Pollmann2009}. The Ansatz can then be understood in terms of the usual finite-size scaling ansatz $\hat O L^{\beta/\nu} = f(\delta J_2 L^{1/\nu})$ with effective length scale $L \sim \xi_\chi \sim \chi^\kappa$. 

\begin{figure}
    \centering
    \includegraphics{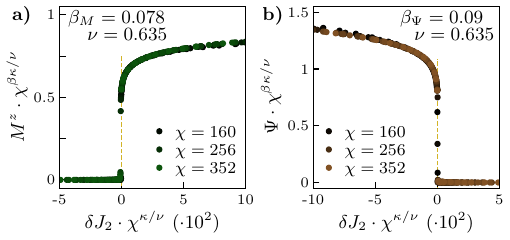}
    \caption{Critical exponents $\beta$ and $\nu$ obtained from a sweep along the frequency direction at the critical endpoint $\omega_c=0.45J_1$ discussed in the main text. The resulting values are consistent with those obtained from the sweep in $J_2$-direction shown in Fig.~\ref{fig:4potts-finite-size}.}
    \label{fig:4potts_finite_size_omega_direction}
\end{figure}

We confirm the critical exponents $\beta$ and $\nu$ in Fig.~\ref{fig:4potts_finite_size_omega_direction} for a sweep \emph{along the frequency direction}, keeping $J_{2c}(\omega_c=0.45J_1)=0.14546J_1$ fixed and varying $\omega_0/J_1$. We obtain $\beta_M = 0.078$, $\beta_\Psi=0.09$ and $\nu=0.635$, in good agreement with previous estimates in Fig.~\ref{fig:4potts-finite-size} and the exact 4-state Potts values $\beta=1/12$, $\nu=2/3$. The result confirms that the four-Potts critical exponents are found independent of the chosen sweep direction.

\end{document}